\documentclass[epj]{svjour}
\usepackage{graphicx}
\usepackage{color}
\usepackage{esint}
\usepackage{bm}
\usepackage{amssymb}

\newcommand{\avg}[1]{\left< #1 \right>} 
\newcommand{\ket}[1]{\left| #1 \right>} 
\newcommand{\bra}[1]{\left< #1 \right|} 

\newenvironment{sistema}%
{\left\lbrace\begin{array}{@{}l@{}}}%
{\end{array}\right.}

\def\rank{\mathop{\textrm{rank}}\nolimits}
\def\tr{\mathop{\textrm{tr}}\nolimits}

\begin{document}
\title{TYPICAL ENTANGLEMENT}
\author{FABIO DEELAN CUNDEN\inst{1} 
\and PAOLO FACCHI\inst{2}
\and GIUSEPPE FLORIO\inst{3,2}
\and SAVERIO PASCAZIO\inst{2}
}                     
%
%
\institute{Dipartimento di Matematica, Universit\`a
di Bari, I-70125 Bari, Italy
\and
Dipartimento di Fisica and MECENAS, Universit\`a
di Bari and  INFN Sezione di Bari, I-70126 Bari, Italy
\and 
Museo Storico della Fisica e Centro Studi e Ricerche
``Enrico Fermi'', I-00184 Roma, Italy
}
\date{Received: date / Revised version: date}
%
\abstract{
Let a pure state  $\ket{\psi}$ be chosen randomly in an $NM$-dimensional Hilbert space, and consider the reduced density matrix $\rho_A$ of an $N$-dimensional subsystem. The bipartite entanglement properties of $\ket{\psi}$ are encoded in the spectrum of $\rho_A$. By means of a saddle point method  and using a ``Coulomb gas'' model for the eigenvalues, we obtain the typical spectrum of reduced density matrices. We consider the cases of an unbiased ensemble of pure states and of a fixed value of the purity. We finally obtain the eigenvalue distribution by using a statistical mechanics approach based on the introduction of a partition function.
} 
\maketitle

\section{Introduction}
\label{intro}

In the last years, many efforts have been directed towards the study of random quantum correlations  \cite{wootters,hall}. This work is a short overview of some important results on the distribution of entanglement among two subparts of a large quantum system. This has been a subject of interest among physicists and mathematicians for a long time, and many interesting results have been achieved, for instance in the context of quantum maps  \cite{scott,rossini}. Our presentation has the style of a pedagogical review, in the sense that many important results will often be rederived by simpler methods. The mathematical techniques will be easy to follow, and will often be guided by physical insight. 

Our aim is to find the maximum of the joint distribution of the eigenvalues of the reduced density matrix. We shall therefore focus on the most probable spectrum of the reduced density matrices. In order to solve the saddle point equations, where the gradient vanishes,  
we shall invoke the physical interpretation of the so-called ``Coulomb gas''  \cite{Dyson}, displaying the profound link between our problem and constrained $2D$-electrostatic models. It turns out that the most probable eigenvalues of the density matrix coincide with the equilibrium positions of movable charges on a line when the interaction forces arise from a logarithmic potential. It is very fascinating to discover that this equilibrium configuration (the $N$-tuple of eigenvalues that maximize probability) is reached at the zeros of a class of orthogonal 
polynomials. Therefore, the spectrum of a typical quantum state is completely determined by the zeros of a certain polynomial. Even for high-degree polynomials, we shall obtain some useful analytic results (in a compact and manageable form) for some entanglement quantifiers, such as purity. 

All these results agree with known results about the statistical averages of typical states obtained with other methods. We shall also  recover many thermodynamic limiting results, generally obtained in literature with methods based on statistical mechanics or random matrix theory  \cite{FacchiPascazio,ADePasquale,ADePasquale2,majumdar,vivo} (for a review of random matrix theory, see Ref.  \cite{mehta}). We will try to avoid unnecessary  mathematical details or unclear physical hypotheses.

\subsection{Notation: setting up the problem}

Let us consider a bipartite quantum system $S=A+B$ whose associated Hilbert space is a tensor product
\begin{equation}
\mathcal{H}_S =\mathcal{H}_A\otimes\mathcal{H}_B,\qquad  \textrm{with}\quad \dim{\mathcal{H}_A} = N \leq \dim{\mathcal{H}_B}= M. 
\end{equation}
If the global system $S$ is described by a pure state, that is a unit vector $\ket{\psi}\in\mathcal{H}_S$,  subsystem $A$ is described by the 
reduced density matrix, obtained by tracing out subsystem $B$
\begin{equation}
\rho_A=\tr_B{\ket{\psi}\bra{\psi}}\ ,
\label{eq:1}
\end{equation}
which is known to be a Hermitian, positive, unit-trace $N\times N$ matrix. In terms of its eigenvalues, the spectrum  of $\rho_A$ is a probability vector 
$\bm{\lambda}=(\lambda_1,\dots,\lambda_N)$, formed by $N$ nonnegative numbers $\lambda_j\geq 0$ that sum up to $1$,  i.e.\ $\sum\lambda_j =1$.
The pure quantum states $\ket{\psi}$ of a bipartite system $\mathcal{H}_A\otimes\mathcal{H}_B$ are said to be \emph{separable} if they admit the product form $\ket{\psi}=\ket{\phi}_A\otimes\ket{\chi}_B$ for some $\ket{\phi}_A, \ket{\chi}_B$ belonging to $\mathcal{H}_A$ and $\mathcal{H}_B$, respectively. If the state does not admit such a factorization, it is said to be \emph{entangled}. 

Bipartite pure states represent one of the few cases for which the problem of revealing quantum correlations admits an exhaustive answer. From the Schmidt decomposition, one has to look at the spectrum of the reduced density matrix $\rho_A$; if its spectrum is (a permutation of) $\bm{\lambda} =(1,0,\dots,0)$, that is $\rho_A^2=\rho_A$ is a rank one projection, then the state $\ket{\psi}$ is separable. Otherwise, if $\rank \rho_A >1$, the state $\ket{\psi}$ is entangled. 

A first amount of information about the separability of a state with respect to a given bipartition is provided by the so-called Schmidt number  \cite{schmidtnumber}, defined as the rank of the reduced density matrices of its subparts. In order to better quantify entanglement, one can recall the statistical interpretation of the reduced density matrix: the more entangled the global pure state, the more mixed the reduced state. The eigenvalues of the reduced density matrix form a probability vector and the degree of mixedness of a state is related to how ``nearly equal'' the eigenvalues are. To this end, it is useful to have in mind the basic ideas of majorization theory  \cite{Bhatia}. We recall that the probability vector $\bm{x}$ is said to be majorized by the probability vector $\bm{y}$, in symbols $\bm{x}  \prec \bm{y}$, if 
\begin{equation}
\max_\sigma \sum_{j=1}^k x_{\sigma(j)} \leq \max_\sigma \sum_{j=1}^k y_{\sigma(j)}, \qquad  1\leq k\leq N,
\end{equation}
where $\sigma$ is a permutation. Thus, the eigenvalues of density matrices satisfy
\begin{equation}
(1/N,\dots,1/N) \prec (\lambda_1, \dots, \lambda_N) \prec (1,0, \dots, 0),
\end{equation}
in agreement with the degree of mixedness, and the corresponding degree of entanglement.

A manageable measure of entanglement is the local \emph{purity}: Given a bipartite pure state $\ket{\psi}$ with reduced density matrices $\rho_A$ and $\rho_B$, one defines
\begin{equation}
\pi_{AB}=\tr_A\rho_A^2=\tr_B\rho_B^2=\sum_{j=1}^N {\lambda_j^2},\qquad 1/N\leq\pi_{AB}\leq 1\ .
\label{eq:purity}
\end{equation}
The local purity 
 has the Schur-convexity property, i.e., it preserves majorization order: 
 \begin{equation}
\pi_{AB} (\bm{\lambda}) \leq \pi_{AB}(\bm{\mu}), \qquad  \textrm{if} \quad  \bm{\lambda} \prec \bm{\mu}.
\end{equation} 
As such, it is a good entanglement measure. In particular $\pi_{AB}=1$ and $\pi_{AB}=1/N$ iff $\ket{\psi}$ is, respectively, separable and maximally entangled with respect to the given bipartition. 

Let us start to discuss the typical properties of random pure states, that is unit vectors  ``drawn at random'' from the Hilbert space.  The rationale behind the sampling criterion is to   introduce no bias in our judgments, namely information that we do not have. The idea of randomly taking a pure quantum state is then equivalent to assuming minimal a priori knowledge about the system. Identifying minimal knowledge with maximal symmetry, it is natural to require that the statistical ensemble be invariant under the full group of unitary transformations. Thus the sampling criterion corresponds to a unique ``natural'' measure on states, induced by the Haar probability measure $d\mu_H(U)$ on the unitary group $\mathcal{U}(L)$, where $L= M N = \dim \mathcal{H}_S$. In other words, a random pure state, defined by the action $\ket{\psi} = U\ket{\psi_0}$ of a random unitary matrix $U$ on a given reference state $\ket{\psi_0}$, can be represented, in an arbitrary basis, as a given column of the random unitary matrix $U$.
Such an ensemble may be identified as the ``most random'' ensemble of possible states of the system.
By tracing over subsystem $B$, this measure translates into a measure over the space of Hermitian, positive matrices of unit trace  \cite{ZykzSommers}. 
By standard methods 
one can show that the measure $d \mu(\rho)$ of density matrices $\rho\in\mathcal{D}(N)$   factorizes into a product measure, with respect to the diagonalization $\rho = U^\dagger \mathrm{diag}(\bm{\lambda}) U$, where $U\in \mathcal{U}(N)$ and $\mathrm{diag}(\bm{\lambda})$ is the diagonal matrix with diagonal entries $\bm{\lambda}$, 
\begin{equation}
d\mu(\rho)=d\nu(\bm{\lambda}) \times d\mu_H (U)\ ,
\end{equation}
where the first factor defines a measure on the $(N-1)$-dimensional simplex of the probability vectors of eigenvalues 
\begin{equation}
\Delta_{N-1} = \{ \bm{\lambda}\in \mathbb{R}^N \,|\, \lambda_k\ge0, \sum_k \lambda_k=1\}. 
\end{equation}
The second factor $\mu_H$ on the space of unitary matrices $\mathcal{U}(N)$ is responsible for the choice of the eigenvectors of $\rho$. A unitarily invariant measure over the pure states of a composite system induces a measure over the eigenvectors of the reduced density matrix $\rho_A$ which is still rotationally invariant, i.e.\ a Haar measure: Think for example of the measure on an equator induced by a uniform measure on a hypersphere. Observe that the space of Hermitian matrices 
is not compact, while the space 
of states $\mathcal{D}(N)$  is the  product of two compact spaces, a simplex and a sphere.

\section{Joint distribution of the eigenvalues} 

\label{sec:joint}

Since the information about the separability of a bipartite pure state $\ket{\psi}$ is completely encoded in the spectrum of its reduced density matrix $\rho_A$, we will focus our attention on the typical properties of the eigenvalues $\bm{\lambda}$ of $\rho_A$.
For random pure states sampled from the unbiased ensemble $\Bigl(\left\{U\ket{\psi_0}\bra{\psi_0}U^{\dagger}\right\}_{U\in\mathcal{U}(L)},d\mu_{H}(U)\Bigr)$,
the eigenvalues of the reduced $N-$dimensional density matrix $\rho_A=\tr_B(U\ket{\psi_0}\bra{\psi_0}U^{\dagger})$
are distributed according to the measure $d\nu(\bm{\lambda})= f_{N,M}(\bm{\lambda}) d \bm{\lambda}$, with joint probability density function (pdf)  \cite{LloydPagels,Page}
\begin{equation}
f_{N,M}(\bm{\lambda})=C_{N,M}\prod_{1\leq j<k\leq N}{(\lambda_j-\lambda_k)^2}\prod_{1\leq l\leq N} \lambda_l^{M-N}
\ ,
\label{eq:Haar_invariant}
\end{equation}
where  \cite{ZykzSommers}
\begin{equation}
C_{N,M}=\frac{(NM-1)!}{\prod_{1\leq j\leq N}{(M-j)!(N-j+1)!}}
\end{equation}
is a normalization factor, assuring that 
$\nu(\Delta_{N-1}) 
= \int_{\Delta_{N-1}} f_{N,M}(\bm{\lambda}) d \bm{\lambda} = 1 .$

Let us summarize some known results about unbiased random states and the eigenvalues distribution~(\ref{eq:Haar_invariant}). 
From the permutation invariance of the eigenvalues joint pdf and the unit trace condition, one immediately obtains:
\begin{equation}
1=\avg{\sum_i{\lambda_i}}=\sum_i{\avg{\lambda_i}}=N\avg{\lambda_i}\quad\Longrightarrow\quad \avg{\lambda_i}=\frac{1}{N},\quad\,\forall\, 1\leq i\leq N\ ,
\label{eq:avg_lambda}
\end{equation}
where $\avg{\cdot}$ stands for the expectation value with respect to the pdf (\ref{eq:Haar_invariant}).
Calculation of the second moment 
needs more work. Lubkin  \cite{Lubkin} calculated 
\begin{equation}
\sigma_{\mathrm{rms}}=\sqrt{\avg{\left(\lambda_i-\frac{1}{N}\right)^2}}=\left(\frac{1-1/N^2}{MN+1}\right)^{1/2}\ .
\label{eq:Lubkin_rms}
\end{equation}
A remarkable fact is that, in order to perform his calculation, Lubkin did not use the above joint pdf but rather a geometric method requiring averages over the real $2NM$-dimensional unit sphere (the joint distribution was discovered many years later  \cite{LloydPagels}).
By denoting $\mu$ s.t. $\mu N=M-N$, in the large $N$ limit, with the ratio $M/N = 1 + \mu$ finite and fixed, the width of the distribution becomes
\begin{equation}
\sigma_{\mathrm{rms}}\sim\frac{1}{N\sqrt{1+\mu}}\ .
\label{eq:thermo_rms}
\end{equation}

Various aspects of the entanglement properties of random pure states have been studied in previous articles. From~(\ref{eq:avg_lambda}), (\ref{eq:Lubkin_rms}) and permutation invariance one obtains the average value of purity
\begin{equation}
\avg{\pi_{AB}}= \avg{\sum_i \lambda_i^2} = N \avg{\lambda_i^2} = N \sigma_{\mathrm{rms}}^2 + N \avg{\lambda_i}^2=
\frac{N+M}{MN+1}\ .
\label{eq:Lubkin}
\end{equation}
For $M/N$ finite and fixed, and large $N$ one has 
\begin{equation}
\avg{\pi_{AB}}\sim \frac{1}{N}\frac{2+\mu}{1+\mu}\ .
\label{eq:purity_unbalanced}
\end{equation}
A balanced $(\mu=0$) bipartite large system has  typical purity $\avg{\pi_{AB}} \sim 2/N$  \cite{FacchiPascazio}.

Another measure of bipartite entanglement for pure states is the local von Neumann entropy $S_{AB}=\tr{\{\rho_A\ln{\rho_A}\}}$. Its average over  random pure states  is given by
\begin{equation}
\avg{S}=\sum_{k=M+1}^{NM}\frac{1}{k}-\frac{N-1}{2M}\ ,
\label{eq:Page}
\end{equation}
as conjectured by Page in his pioneering work  \cite{Page}, and proved in  \cite{FoongKanno}. See also  \cite{Sen}. 

As a final quantity that provides information on the degree of mixedness of $\rho_A$, and thus of the entanglement of a pure state, we recall the elementary symmetric polynomials $s_k(\bm{\lambda})=\sum_{j_1<\cdots<j_k}{\lambda_{j_1}\cdots\lambda_{j_k}}$, with $1<k\leq N$. In particular, the elementary invariant $s_N(\bm{\lambda})=\lambda_1\cdots\lambda_N$ is nothing but the determinant of the density matrix $\det{\rho_A}$. It is a bounded Schur-concave function of the spectrum 
of the reduced density matrix. If the state is sampled according to the unbiased ensemble, the form of the moments of $\det{\rho_A}$ is a straightforward consequence of Eq.~(\ref{eq:Haar_invariant})
\begin{equation}
\avg{\det{\rho_A^k}}_{N,M}=\frac{C_{N,M}}{C_{N,M+k}}\ .
\label{eq:moments_e_N}
\end{equation}

\section{Most probable distribution}
\label{sec:most_probable}

In the previous section we have presented some important results about the expectation values of some interesting entanglement quantifiers. 
An alternative approach relies on the study of the typical properties. The idea is that, given a function of random states $h(\ket{\psi})$, which is therefore itself a random variable, its most probable value $\tilde{h}$ is ``close'' to its average $\avg{h}$. Then, the typical properties of the reduced state $\rho_A$ depend on its typical spectrum.
We look for the most probable eigenvalues, that is  the point(s) $\bm{\lambda}$ that maximizes the pdf~(\ref{eq:Haar_invariant}) on the simplex $\Delta_{N-1}$.
If one writes the function~(\ref{eq:Haar_invariant}) 
as a ``Boltzmann factor''
\begin{equation}
f_{N,M}(\bm{\lambda})=C_{N,M} \exp{\left(-F_{N,M}(\bm{\lambda})\right)}\ ,
\label{eq:Boltzmann}
\end{equation}
then the points that maximize the probability $f$ are points of  minimum for the ``energy'' $F$:
\begin{equation}
F_{N,M}(\bm{\lambda})= - \ln{\frac{f _{N,M}(\bm{\lambda})}{C_{N,M}}}=- 2\sum_{i<j}\ln{|\lambda_i-\lambda_j|} - (M-N)\sum_l{\ln{\lambda_l}}
\label{eq:2}
\end{equation}
on the simplex $\Delta_{N-1}$.

\begin{figure}[t]
\centering
\includegraphics[width=0.23\columnwidth]{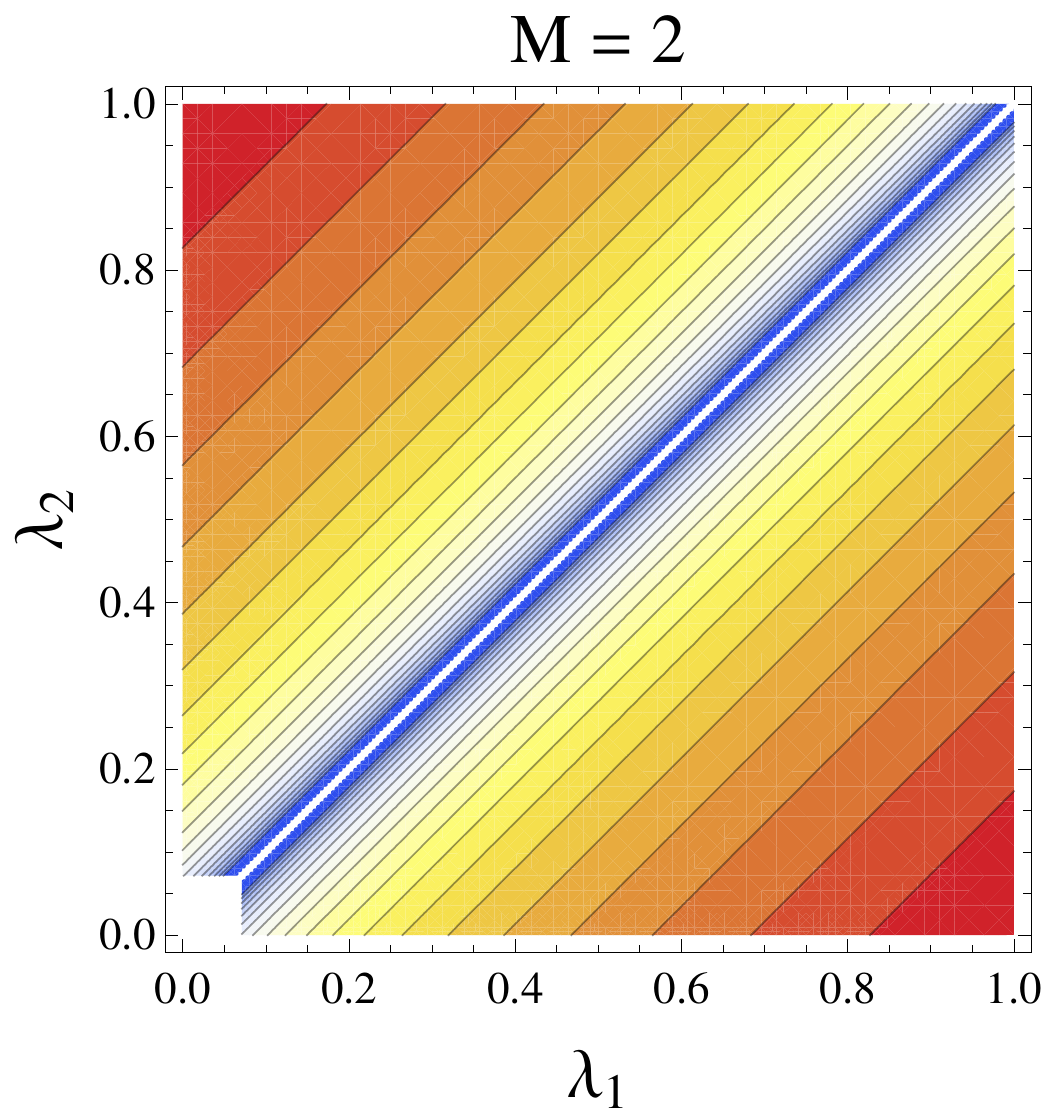}
\includegraphics[width=0.23\columnwidth]{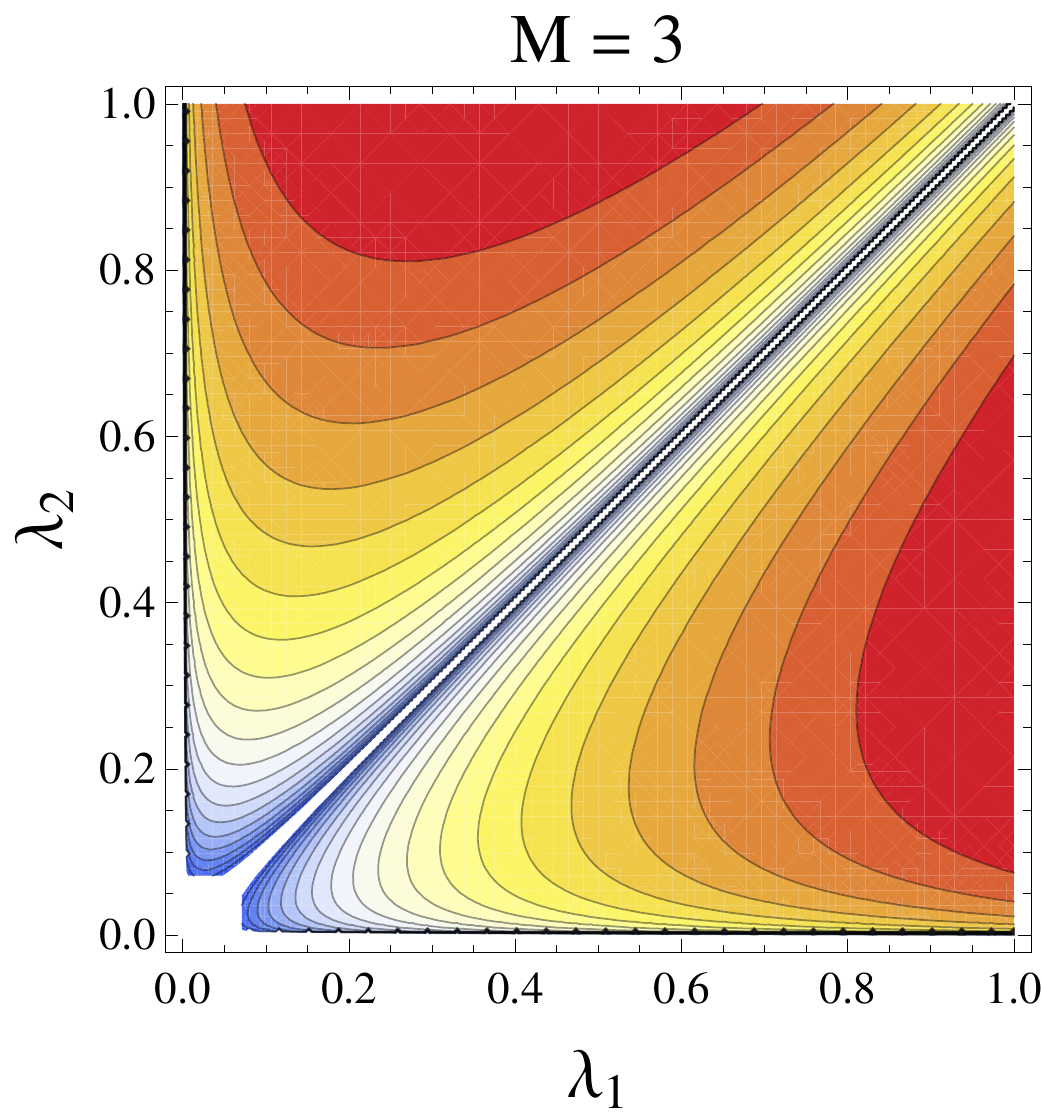}
\includegraphics[width=0.23\columnwidth]{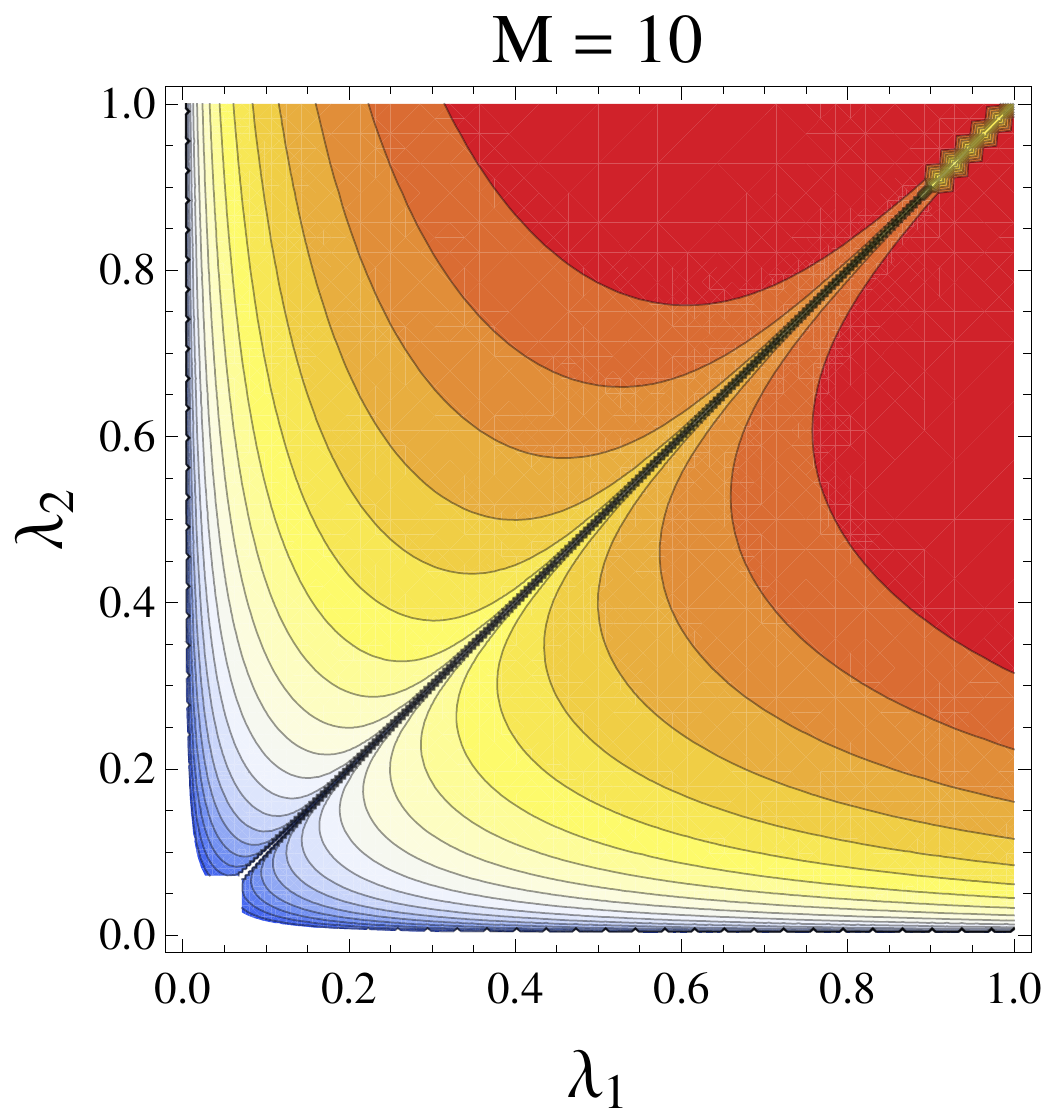}
\includegraphics[width=0.23\columnwidth]{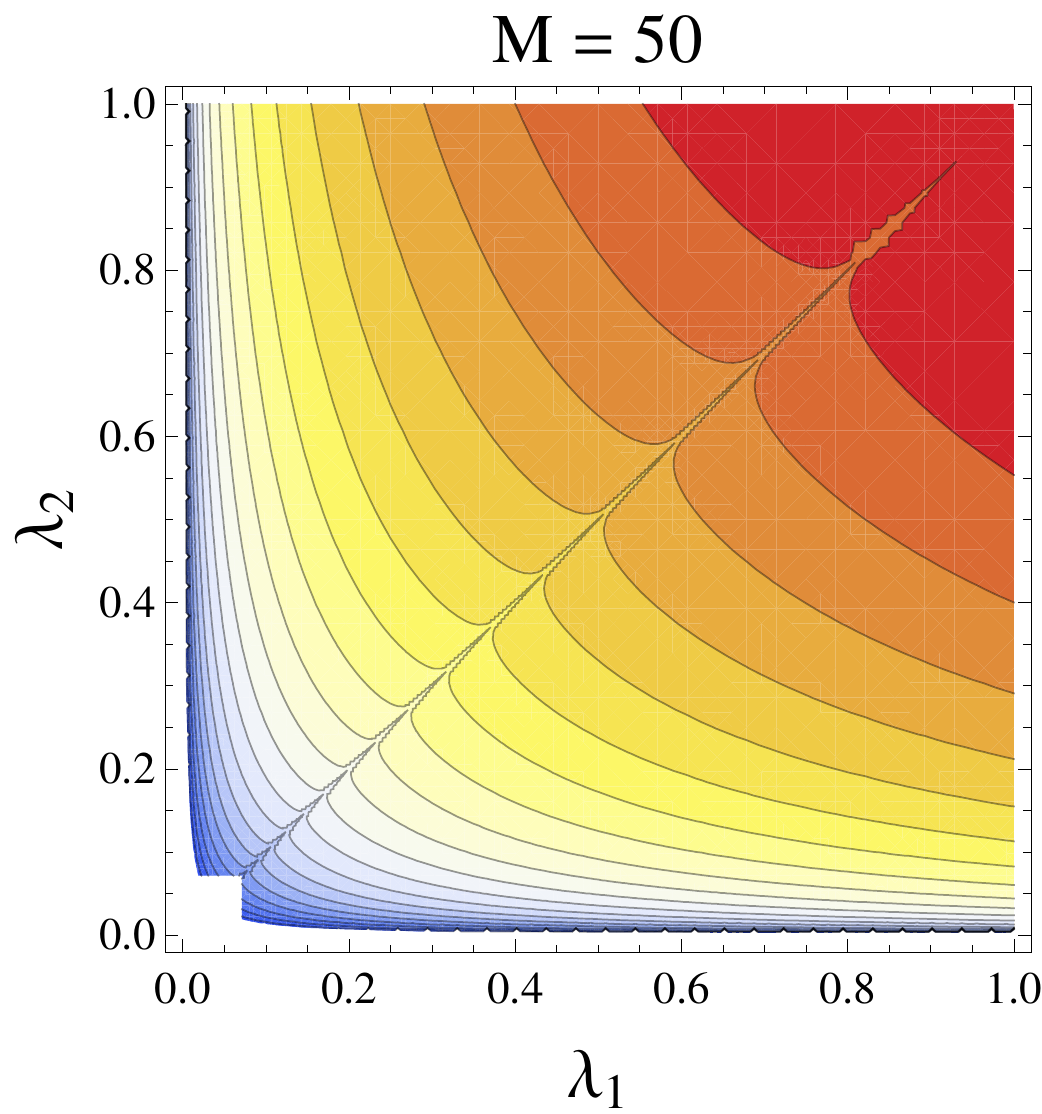}\\
\includegraphics[width=0.23\columnwidth]{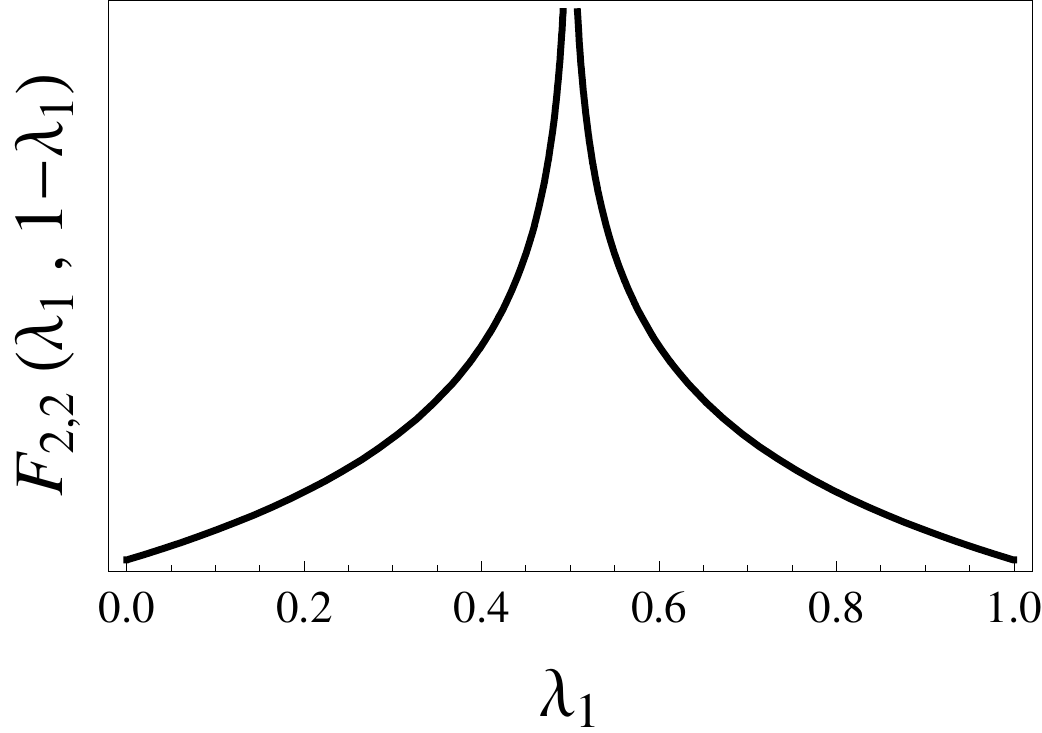}
\includegraphics[width=0.23\columnwidth]{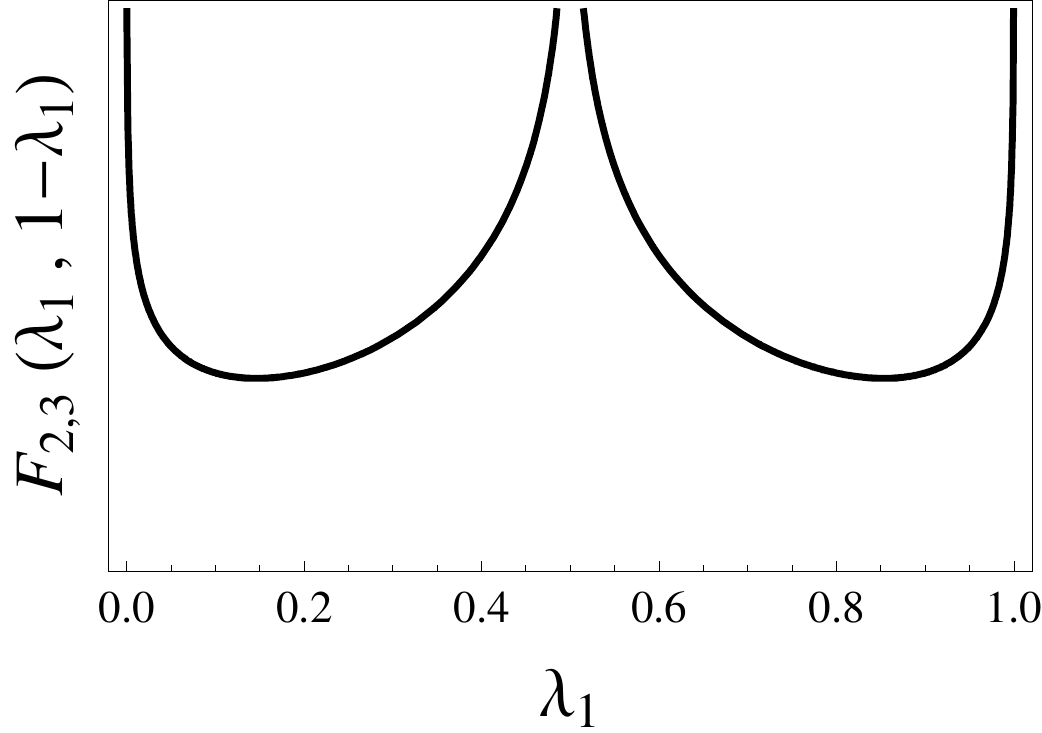}
\includegraphics[width=0.23\columnwidth]{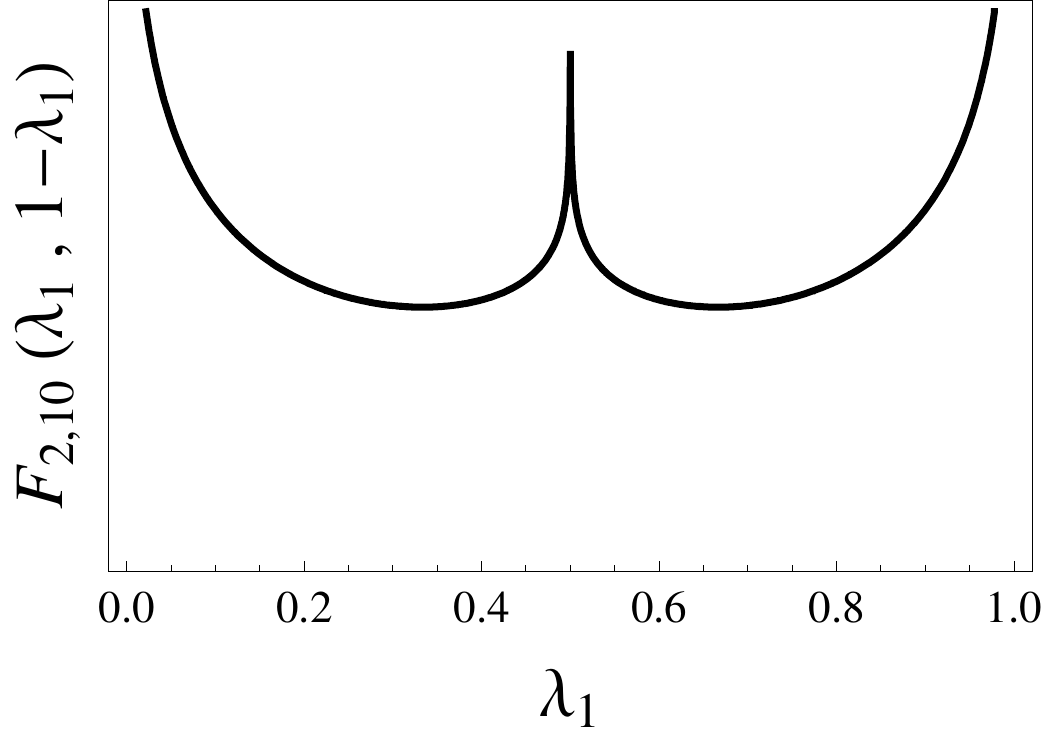}
\includegraphics[width=0.23\columnwidth]{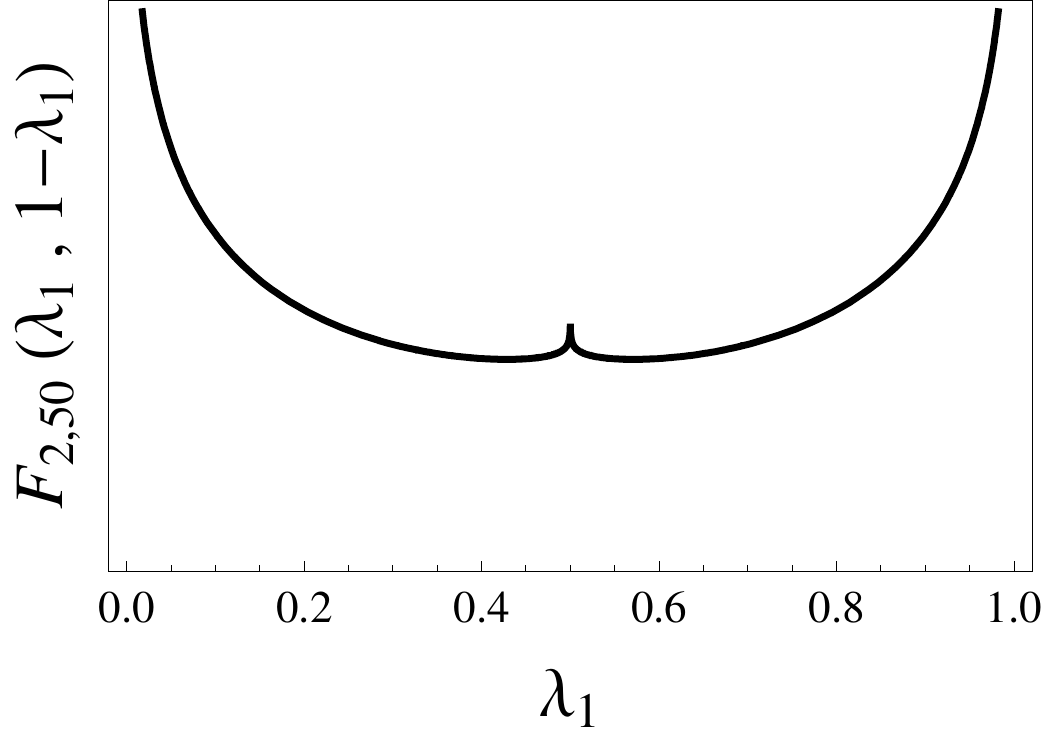}
\caption{Contour plots of the function $F_{2,M}$ for a single qubit ($N=2$) for different dimensions of the environment $M$; the domain is the full square $\lambda_1,\lambda_2\in\left[0,1\right]\times\left[0,1\right]$. Below, the restrictions of $F_{2,M}$ to the simplex (i.e. the diagonal of the square from $(1,0)$ to $(0,1)$ ). When $M>N$, two minima appear in the interior of the simplex. 
As $M$ increases, the repulsion between the eigenvalues becomes dominated by a ``uniforming'' effect of the environment.}
\label{fig:contour}
\end{figure}
The function $F_{N,M}$ is smooth in the subset of the simplex $\Delta_{N-1}$ defined by the inequalities $\left|\lambda_i-\lambda_j\right|>0$ and $\lambda_i>0$ if $M>N$. 
Moreover, when $\lambda_i - \lambda_j \to 0$ for some $i \neq j$ the energy diverges, $F_{N,M}\to +\infty$. When $M>N$ the same thing happens when $\lambda_i \downarrow 0$ for some $i$. Thus, in the unbalanced case there exists a finite minimum of the energy in the interior of the simplex, and the point of minimum is a critical point of the energy, i.e. its gradient vanishes. For $\bm{\lambda} \in \Delta_{N-1}$ one obviously gets $|\lambda_i -\lambda_j|\leq 1$ and $\lambda_i \leq 1$ for any $i,j$, whence $F_{N,M} \geq 0$, and the minimum value is nonnegative.

On the other hand, when $M=N$, the energy function no longer diverges on the boundary, and the minimum energy can be attained on the boundary, and in general it is not a critical value. 
For example, for a qubit, i.e.\ $N=M=2$, the minima are at the boundary points  $(0,1)$ and $(1,0)$, where $F_{2,2}=0$ and $\nabla F_{2,2}\neq 0$ (see Fig. \ref{fig:contour}). In fact, we will show that this is always the case: one eigenvalue, say $\lambda_N$ vanishes at the minimum point, and the problem is reduced to the minimization of $F_{N,N}(\lambda_1, \dots, \lambda_{N-1},0) = F_{N-1,N+1}(\lambda_1, \dots, \lambda_{N-1})$. However, this is nothing but the energy of the unbalanced problem on the simplex $\Delta_{N-2}$, whose minimum is a critical point. 

Summarizing, the problem of maximizing the pdf $f_{N,M}$ reduces to the problem of finding the critical values of  the nonnegative  energy function $F_{N,M}$, when $M>N$ (or $F_{N-1,N+1}$, when $M=N$), in the interior of the simplex $\Delta_{N-1}$ (or $\Delta_{N-2}$).

\section{Unbiased Pure States}

We will start from unbiased states, i.e. states sampled according to the unitarily invariant Haar measure. In particular, we are concerned with the Schmidt coefficients with respect to a given bipartition. These coefficients give information about the degree of mixedness of the reduced density matrices of the subparts of the global system. We are interested in the \emph{typical} entanglement of a small subsystems of a large random pure state. Our approach will rely on a saddle point method: given the joint distribution of the eigenvalues of the reduced density matrix $\rho_A\in\mathcal{D}(N)$, 
we will search the most probable spectrum, that is, the density matrix (up to local unitaries $U\in\mathcal{U}(N))$ that maximizes the probability.

The saddle point problem for typical states can be reduced to the problem of finding the equilibrium configurations of a system of identical movable charges on a line (electrostatic models). These problems are elegantly connected with the theory of orthogonal polynomials, as Stieltjes first showed  \cite{Stieltjes,Szego}.

For unbiased states, one is able to fully solve the problem. The complete solution of the saddle point method will be provided for all possible (unbalanced) bipartitions. This result is the starting point to compute all quantities of interest. For some well-known quantities, such as purity or elementary symmetric invariants, we are able to give compact and manageable analytic expressions, for all $N$ and $M$. Moreover, we will present typical entanglement properties for the unbalanced and balanced bipartition in the large sizes limit $N,M\rightarrow+\infty$. 

\subsection{Coulomb Gas}
\label{sub:heuristic}
\begin{figure}[t]
\centering
\includegraphics[width=0.45\columnwidth]{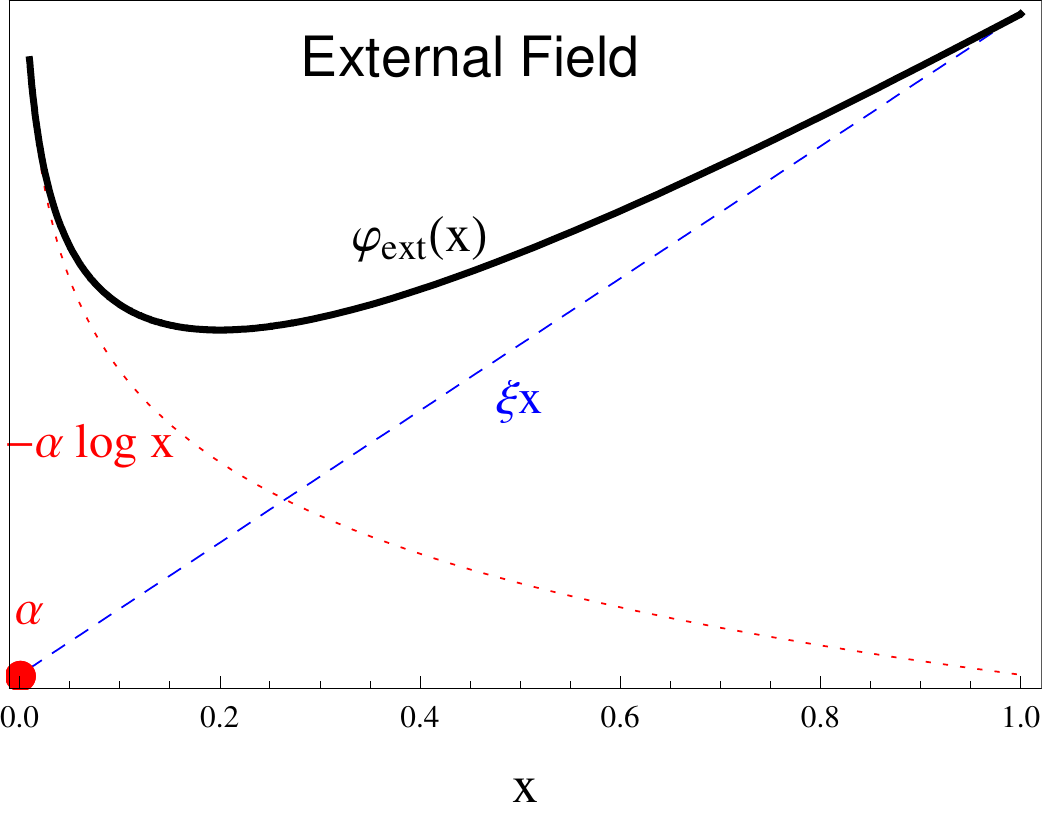}
\includegraphics[width=0.45\columnwidth]{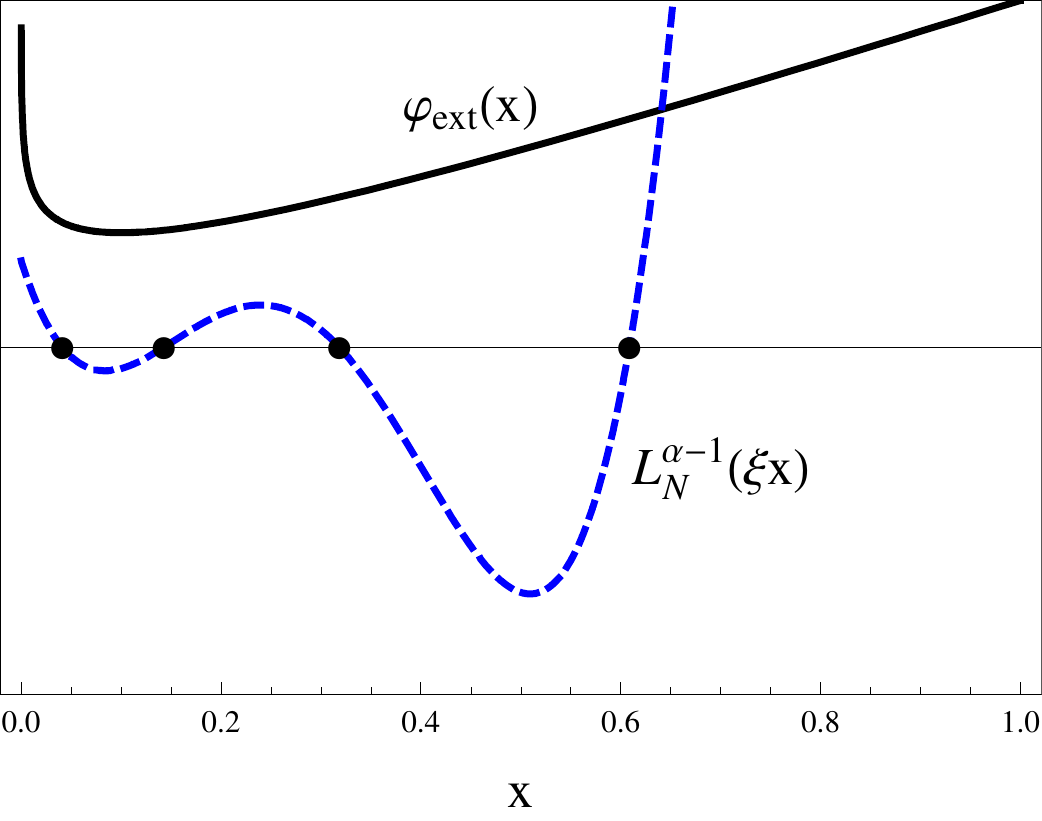}
\caption{Left: Coulomb gas charges experience an external force $-\partial_x\varphi_{\rm{ext}}(x)$. The external potential (\ref{eq:ext_field_mutual_int}) is the sum of a logarithmic part, due to a charge $\alpha=M-N$ at the origin, and a linear part governed by the Lagrange multiplier $\xi$. Right: the equilibrium positions of the charges lie at the zeros of a generalized Laguerre polynomial.}
\label{fig:ExtLaguerre}
\end{figure}

To get a clearer insight of the joint pdf of the eigenvalues, Eq.\ (\ref{eq:Haar_invariant}), one can  invoke the physical picture of a ``Coulomb gas'' of $N$ repelling electric charges on a segment \cite{Dyson}. Indeed, according to the discussion of section~\ref{sec:most_probable}, the most probable distribution is the result  of a constrained minimization problem for an energy function $F_{N,M}$, with suitable $N<M$, that can be handled by using the method of Lagrange multiplier. Namely, one has to find the minima of the $(N+1)$-variable function:
\begin{equation}
E_{\mathrm{tot}}(\bm{\lambda},\xi)=-2\sum_{i<j}\log{|\lambda_i-\lambda_j|}-(M-N)\sum_l{\log{\lambda_l}}-\xi(1-\sum_i{\lambda_i}) ,
\label{eq:landscape_with_constraint}
\end{equation}
that can be interpreted as the potential energy of a gas of $N$ point charges at positions $\lambda_i$'s. The potential energy is given by the mutual repulsion of these charges, plus a part given by an external field
\begin{eqnarray}
V_{\rm{mutual}}(x,y)& =&- 2\ln{|x-y|}\ ,\\
\varphi_{\rm{ext}}(x)& =&-\alpha\ln{x}+\xi x\ ,
\label{eq:ext_field_mutual_int}
\end{eqnarray}
where we have denoted $\alpha=M-N$. The external potential is plotted in Fig.\  \ref{fig:ExtLaguerre}. 
In other words, the total electrostatic  energy is
\begin{equation}
E_{\rm{tot}}(\bm{\lambda}; \xi)= \sum_i \sum_{j\neq i}V_{\rm{mutual}}\left(\lambda_i,\lambda_j\right)+\sum_{i} {\varphi_{\rm{ext}}(\lambda_i)}\ .
\label{eq:total_energy}
\end{equation} 
We are interested in the stationary points of this energy. By  deriving $E_{\rm{tot}}(\bm{\lambda}; \xi)$ with respect to both the $\lambda_i$'s and $\xi$, we get $N+1$ saddle point equations: 
\begin{equation}
\begin{sistema}
{\displaystyle 2\sum_{j\neq i}{\frac{1}{\lambda_i-\lambda_j}}+\frac{M-N}{\lambda_i}-\xi=0\,,\qquad 1\leq i \leq N}\\\\
{\displaystyle \sum_i{\lambda_i}=1\ .}
\end{sistema}
\label{eq:SaddlePointEqns}
\end{equation}
In the framework of the electrostatic model, the saddle point equations are nothing but static equations of balance of the forces (the derivatives of the potential energy, $\nabla_{\bm \lambda}E_{\rm{tot}}=0$), with the additional constraint that the charges average position be equal to $1/N$.

Before trying to write down a solution for the saddle point equation, it is useful to give a picture of what will happen. The $N$ charges interact via a logarithmic 2D-Coulomb repulsion, that is the form of Gauss law in two dimensions. Therefore, in absence of an external field, the charges will move as far apart as possible. However, there are two external forces acting on them. The first one is due to a charge $\alpha$, with $\alpha=M-N>0$, at position $x=0$ that repels the unit charges (the eigenvalues) via a logarithmic potential. This repulsion constrains the charges on the positive half-line. The second field is a constant force (the gradient of the linear potential $\xi x$). In order to ensure the existence of an  equilibrium configuration, the external potential  must have a minimum. Then, $\xi$ must be positive (necessary condition to have a convex potential). 

Notice that in the balanced case, $\alpha=M-N=0$, there is no logarithmic repulsion from the origin. Therefore, one of the $N$ charges will sit at the origin and repel the remaining $N-1$ charges with a potential $-2 \ln x$. This is nothing but the electrostatic problem generated by the energy function $F_{N-1,N+1}$, thus proving that in the balanced case the minima are located on the boundary of the simplex $\Delta_{N-1}$, as discussed in section~\ref{sec:most_probable}.

The minimum of the external potential is located at $x_c= \alpha / \xi$. Since we expect that, in the typical case, all eigenvalues be located near the maximally mixed value, this critical point has to be close to $1/N$. By setting the trial value $x_c\simeq 1/N$, we can immediately guess that $\xi\simeq NM$ in the large-$M$ limit. 
In fact we can do better and compute exactly the Lagrange multiplier by means of a nice trick. Observe that, for any continuous function $h$, the equality 
\begin{equation}
\sum_i \sum_{j\neq i}{\frac{h(\lambda_i)}{\lambda_i-\lambda_j}}=\sum_{i<j}{\frac{h(\lambda_i)-h(\lambda_j)}{\lambda_i-\lambda_j}} 
\label{eq:dirty_trick}
\end{equation} 
holds. 
Consider now the first $N$ saddle point equations~(\ref{eq:SaddlePointEqns}) and multiply each of them by $\lambda_i$ to obtain
\begin{equation}
2\sum_{j\neq i}{\frac{\lambda_i}{\lambda_i-\lambda_j}}+(M-N)-\xi\lambda_i=0\ ,\qquad 1\leq i\leq N .
\label{eq:3}
\end{equation}  
By summing over $i$, and using~(\ref{eq:dirty_trick}) and  the unit-trace condition, one  finds 
\begin{equation}
N(N-1)+N(M-N)-\xi =0\ ,
\label{eq:4}
\end{equation}
whence the sought multiplier reads 
\begin{equation}
\xi=N(M-1) ,
\label{eq:lag_crash}
\end{equation}
as expected by the above argument. 

Once the Lagrange multiplier is known, we can obtain the typical purity $\pi_{AB}$ with the same trick  by multiplying the first $N$ saddle point equations~(\ref{eq:SaddlePointEqns}) by $\lambda_i^2$ and summing over $i$. Since 
\begin{equation}
2\sum_{i<j}{\frac{\lambda_i^2-\lambda_j^2}{\lambda_i-\lambda_j}}=2\sum_{i<j}(\lambda_i + \lambda_j)= 2(N-1)\sum_i{\lambda_i} = 2(N-1), 
\end{equation}
we get
\begin{equation}
2(N-1) + (M-N) - \xi \pi_{AB}= 0,
\end{equation}
whence
\begin{equation}
\pi_{AB}=\frac{N+M-2}{N(M-1)}\ .
\label{eq:purity_crash}
\end{equation}

As a check that the stationary points of the energy $E_{\rm{tot}}(\bm{\lambda};\xi)$ are in fact minima, one can look at the Hessian matrix
\begin{equation}
\bm{H}(E_{\rm{tot}})=\left(\frac{\partial^2}{\partial\lambda_i \partial\lambda_j}E_{\rm{tot}}(\bm{\lambda};\xi)\right)\ .
\end{equation}
It is easy to see that
\begin{equation}
H_{ii}=2\sum_{j\neq i}\frac{1}{(\lambda_i-\lambda_j)^2}+\frac{M-N}{\lambda_i^2}\,,\qquad H_{ij}=-\frac{1}{(\lambda_i-\lambda_j)^2}\,,\qquad 1\leq i,j\leq N\ .
\label{eq:5}
\end{equation}
Thus the Hessian $\bm{H}$ is a strictly diagonally dominant  symmetric  matrix  with positive diagonal elements. Therefore, it is positive definite everywhere,  and so every stationary point of Eq.\ (\ref{eq:total_energy}) is a local minimum. 

\subsection{The Solution}

The saddle point equations~(\ref{eq:SaddlePointEqns}) can be tackled by using an ingenious method due to Stieltjes, that deals with the electrostatic interpretation
of the zeros of some families of orthogonal polynomials.
The first $N$ saddle point equations~(\ref{eq:SaddlePointEqns})  are ``equivalent'' to a single polynomial equation. Indeed, one can write
\begin{equation}
2\sum_{j\neq i}{\frac{1}{\lambda_i-\lambda_j}}= \frac{g''(\lambda_i)}{g'(\lambda_i)},
\label{eq:Stieltjes}
\end{equation}
where 
\begin{equation}
g(x)=\prod_{k}{\left(x-\lambda_k\right)}
\label{eq:nodal}
\end{equation} 
 is the nodal polynomial whose zeros are  the $\lambda_k$'s.
The above identity is commonly known as  Stieltjes's trick. It can be easily derived by noting that 
\begin{equation}
g'(x) = \prod_k (x-\lambda_k) \sum_{i} \frac{1}{x-\lambda_i}, \qquad g''(x) = \prod_k (x-\lambda_k) \sum_{i} \frac{1}{x-\lambda_i}\sum_{j\neq i} \frac{1}{x-\lambda_j},
\end{equation}
for  $x$ different from every $\lambda_i$, and by taking the limit $x\to \lambda_i$.

Thus, the saddle point equations
can be written as
\begin{equation}
\lambda_i g''(\lambda_i) +(\alpha-\xi \lambda_i)g'(\lambda_i)=0\ ,\qquad 1\leq i\leq N\ .
\label{eq:6}
\end{equation}
The above equations mean that the $N$-degree polynomial  
\begin{equation}
x g''(x) +(\alpha -\xi x)g'(x)
\end{equation}
has its $N$ zeros at $(\lambda_1,\dots,\lambda_N)$, and then it is a multiple of $g(x)$. Comparing the leading coefficients of the two polynomials we get that the proportionality constant is $-N \xi$, and we arrive at the differential equation  
\begin{equation}
x g''(x)+(\alpha -\xi x)g'(x)+\xi N g(x)=0\ ,
\label{eq:ODEcompl}
\end{equation}
where we recall that $g(x)$ is a polynomial whose simple zeros are the unknown $\lambda_i$'s. The polynomial solution of the previous ordinary differential equation is an associated Laguerre polynomial \cite{Szego}
\begin{equation}
L_{N}^{\left(\alpha-1\right)}\left(\xi x\right)= \sum_{\nu=0}^N c_{\nu} (-x)^\nu, \qquad
c_{\nu} = \frac{\xi^{\nu}}{\nu!} {M-1 \choose N-\nu}, \qquad
\alpha=M-N\ .
\label{eq:solLaguerre}
\end{equation}
See Fig.\ \ref{fig:ExtLaguerre}. The value of the Lagrange multiplier $\xi$ is fixed by the trace condition. Since the coefficients of $x^{N-1}$ and $x^N$ are related by $c_{N-1} = c_N \sum_i \lambda_i$,  one should have $c_{N-1}=c_N$, that reads $(M-1)\xi^{N-1}/(N-1)! = \xi^{N}/N!$, yielding 
Eq.~(\ref{eq:lag_crash}). 

Notice finally  that the Lagrange multiplier $\xi$ is related to the trace of $\rho_A^{-1}$. Indeed by just summing the saddle point equations and  using~(\ref{eq:dirty_trick}) with $h=1$, one easily obtain $\xi=\mathrm{Tr}{\rho_A^{-1}}(M-N)/N$ and then $\mathrm{Tr}{\rho_A^{-1}}=N^2(M-1)/(M-N)$.
\begin{figure}[t]
\centering
\includegraphics[width=0.45\columnwidth]{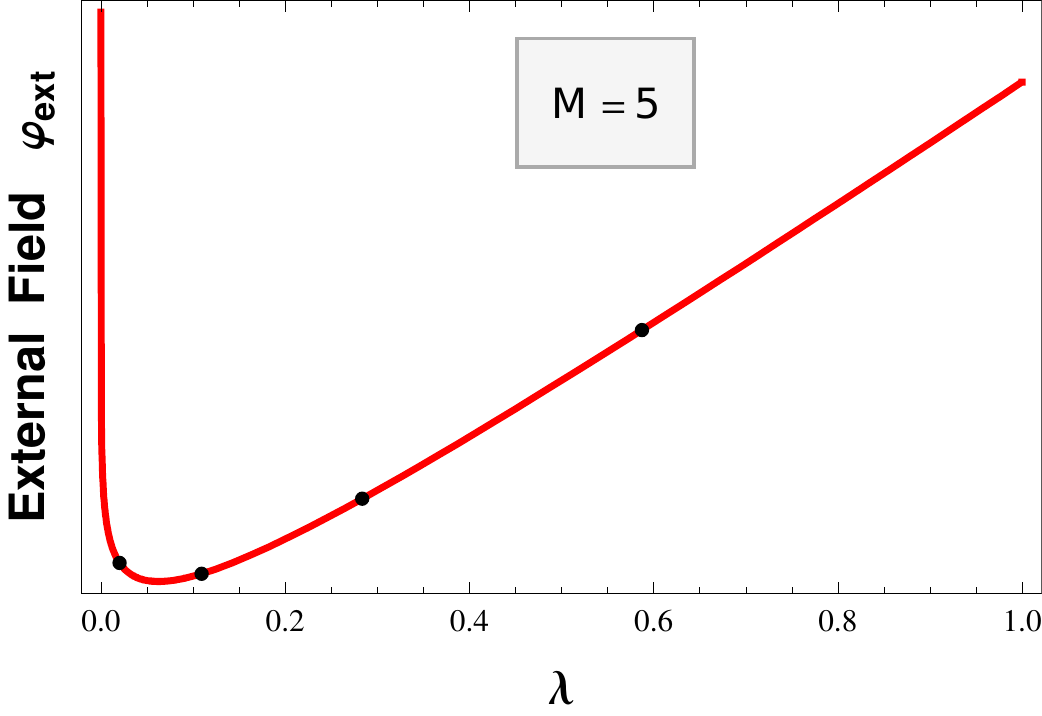}
\includegraphics[width=0.45\columnwidth]{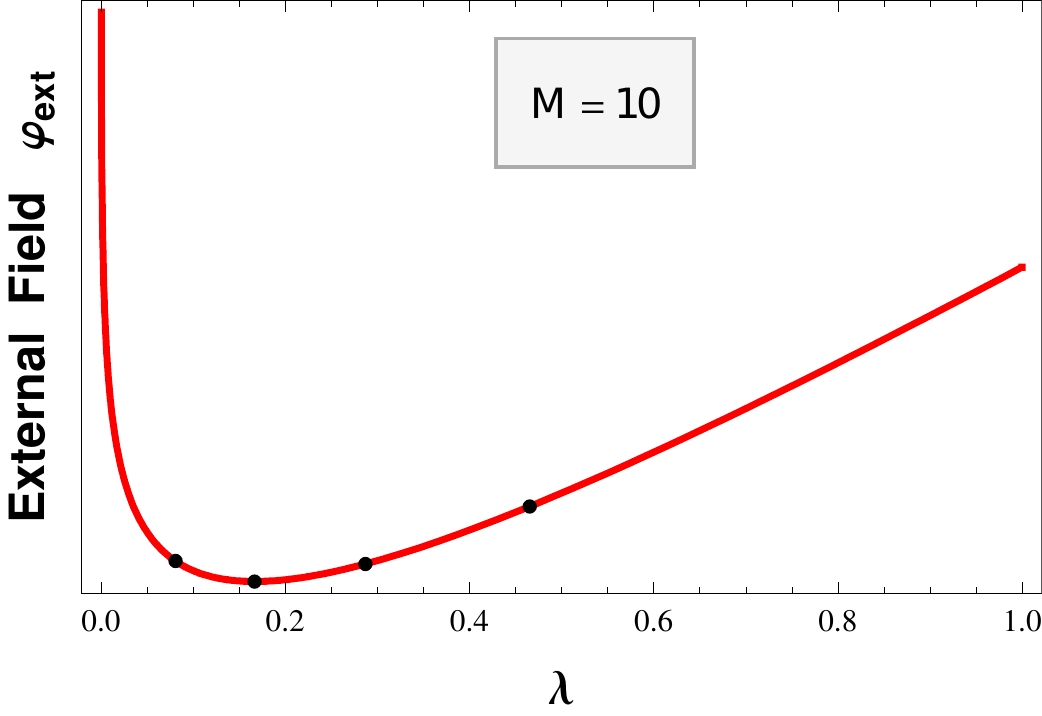}\\
\includegraphics[width=0.45\columnwidth]{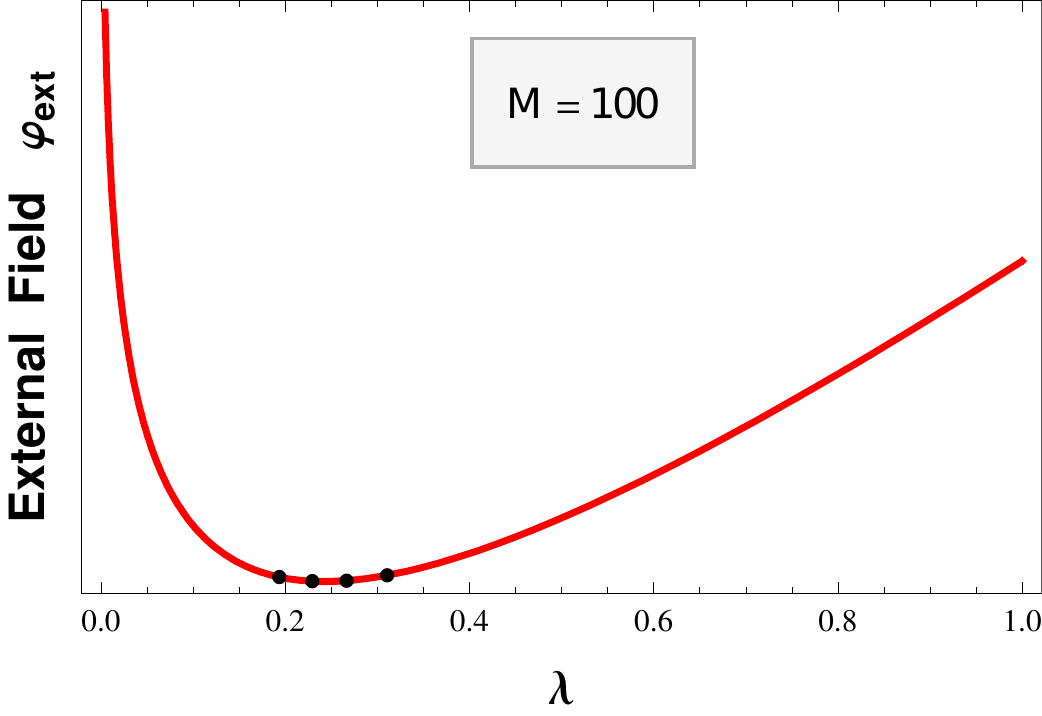}
\includegraphics[width=0.45\columnwidth]{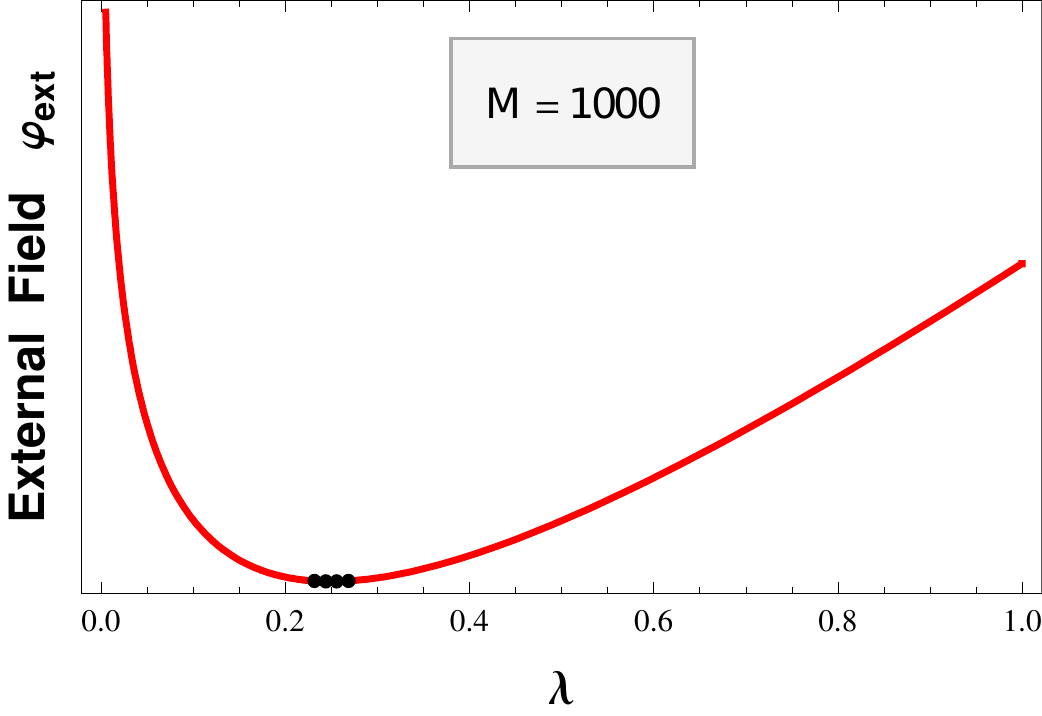}
\caption{Most probable eigenvalues of a four-level system $N=4$ when $\dim{\mathcal{H}_B}=M>4$ varies. In the Coulomb gas framework, the four charges reach their equilibrium position in the external field $\varphi_{\textup{ext}}=-\alpha\ln{x}+\xi x$, where $\alpha=M-4$ and $\xi=4(M-1)$, see Eq.~(\ref{eq:lag_crash}). As $M$ increases the state tends toward the maximally mixed one, i.e.\ $\bm{\lambda} = (1/4,1/4,1/4,1/4)$. }
\label{fig:EquilibriumLaguerre}
\end{figure}

\subsection{Typical entanglement spectrum}
\label{sec:features}

Once the exact solution is found one can study any quantity of interest.
Using general results of elementary algebra, one can extract plenty of information from the coefficients of the Laguerre polynomial. 
For example, in order to compute the purity, Eq.\ (\ref{eq:purity}), note that $\Bigl(\sum_i{\lambda_i}\Bigr)^2=\sum_i{{\lambda_i}^2}+2\sum_{i < j}{\lambda_i\lambda_j}$, that is 
\begin{equation}
\pi_{AB}
=1-2\frac{c_{N-2}}{c_N}=1-\frac{(N-1)(M-2)}{N(M-1)}=\frac{N+M-2}{N(M-1)}, 
\label{eq:purity_typical}
\end{equation}
which is exactly the same as Eq.\ (\ref{eq:purity_crash}), to be compared with the average value computed by Lubkin, Eq.\ (\ref{eq:Lubkin}). 
For large $N$, $\pi_{AB}=O(1/M)$ while the difference between mean and mode is $O(1/NM)$.
Similarly, one can readily compute the elementary symmetric invariants
\begin{eqnarray}
s_k(\bm{\lambda})&=&\sum_{j_1<j_2<\dots<j_k}{\lambda_{j_1}\lambda_{j_2}\dots \lambda_{j_k}}= \frac{c_{N-k}}{c_N}
\nonumber\\
&=& \frac{N!(M-1)!}{k!(N-k)!(M-k-1)!}\left[N(M-1)\right]^{-k}\ .
\label{eq:symmetric_typical}
\end{eqnarray}
As an example, the typical determinant of the reduced state $\rho_A$ is
\begin{equation}
s_N(\bm{\lambda})=\prod_i{{\lambda}_i}=\det{({\rho}_A)}=\frac{(M-1)!}{(M-N-1)!}\left[N(M-1)\right]^{-N}\ .
\label{eq:det_typical}
\end{equation}

Let us conclude this section with a remark. A given property is said to be ``typical'' if it holds with overwhelming probability. In the previous section we just found the most probable spectrum of the reduced state $\rho_A$. The next step whould be to show that the probability concentrates around the most probable value. Some authors \cite{Winter,Mueller} achieved some exact results on the concentration of the spectrum of the reduced density matrix when both subsystem are large. We will give here a heuristic justification of this assertion based on a ``second moment bound''. In what follow we will specialize our discussion to highly unbalanced bipartitions and a large-size limit.
Let us suppose that subsystem $A$ is very small compared with $B$ so that we can think at $B$ as an environment. 
One expects that when  the dimension of the environment $M$  is large, the eigenvalues $\lambda_i$ fluctuate around $1/N$. Thus we make the educated guess
\begin{equation}
\lambda_i=
\frac{1}{N}+\frac{\epsilon_i}{M^{\alpha}}\ .
\end{equation}The value of the variance of the random variable $\lambda_i$ in the unbiased case,
Eq.~(\ref{eq:Lubkin_rms}), suggests that $\alpha=1/2$,
and the saddle point equations in terms of the $\epsilon_i$'s read, for large values of $M$
\begin{equation}
\begin{sistema}
{\displaystyle 2\sum_{j\neq i}{\frac{1}{\epsilon_i-\epsilon_j}}+N\sqrt{M}-N^2\epsilon_i-\xi\sqrt{M}=0\ ,\quad\forall i\in\{1,\dots,N\}}\\\\
{\displaystyle \sum_i{\epsilon_i}=0.}
\end{sistema}
\label{eq:simpli_eq}
\end{equation}
We can use Stieltjes's trick to handle the above simplified saddle point equations. In this framework, the first $N$ equations in (\ref{eq:simpli_eq}) are equivalent to the single differential equation:
\begin{equation}
g''(x)-(N^2 x-N\sqrt{M}+{\xi}\sqrt{M})g'(x)+N^3g(x)=0\ .
\label{eq:ODEsimpl}
\end{equation}
whose polynomial solution $H_N$ is the Hermite polynomial of degree $N$  \cite{Szego}. 
Therefore, the most probable eigenvalues satisfy 
\begin{equation}
H_N\left(\frac{N \epsilon_i}{\sqrt{2}}\right)=0\ .
\label{eq:general_sol}
\end{equation}
This approximated solution is useful to find a second-moment bound with a Gaussian approximation.
The most probable value is typical (and close to the average) if the pdf is sharply peaked. 
The width of the peak is encoded in the Hessian $\bm{H}(\bm{\lambda})$ of the energy, Eq. (\ref{eq:total_energy}), evaluated at the minima. In the Gaussian approximation, the proper values $h_i$'s of $\bm{H}$ give the widths ($\sigma_i^{-2}=h_i$) of the pdf around its maximum.
In order to extract the correct scaling, we can consider just the trace of the Hessian and use the solution of the simplified saddle point equation, Eq.\ (\ref{eq:general_sol}), to obtain
\begin{equation}
\tr\bm{H}\leq N^3(M-N)+2N(N-1)M\ .
\label{eq:hessian_trace}
\end{equation}
Since $\tr\bm{H}=\sum_i{1/\sigma^2_i}\leq N/\sigma_{\textup{min}}^2$, we readily find an upper bound on the width of the Gaussian.

\subsection{Thermodynamic limit}

In this section we want to use our solution to investigate the large size limit of two macroscopic partitions.
As in section~\ref{sec:joint}, it is convenient to introduce the parameter $\mu$ such that $\mu N=M-N$. When $N,M\rightarrow +\infty$ with $\mu$ fixed and finite, the most probable purity reads
\begin{equation}
{\pi}_{AB}=\frac{M+N-2}{NM-1}=\frac{N(2+\mu)-2}{N^2(1+\mu)-1}\sim
\frac{1}{N}\frac{(2+\mu)}{(1+\mu)}\ , \qquad N\to\infty,
\label{eq:purity_therm_lim2}
\end{equation}
a value obtained in~\cite{FacchiPascazio}
by more sophisticated methods. 

Similarly, one can compute the quantities $\tr\rho_A^k$, and then derive the Renyi's entropies. A very inexpensive method is to write $\sum_i{\lambda_i^k}$ in terms of elementary symmetric polynomials, and then use Eq. (\ref{eq:symmetric_typical}). The above procedure can be easily implemented by the most common symbolic manipulating softwares (\emph{Mathematica} 
 provides the suitable function \texttt{SymmetricReduction[]} for expanding any symmetric polynomial). We give a list of the first five traces in the thermodinamic limit, $N,M\rightarrow+\infty$ with $\mu$ fixed: 
\begin{eqnarray}
\tr{\rho_A^2}&=&\frac{(2+\mu)}{(1+\mu)}\frac{1}{N}+O\left(\frac{1}{N^2}\right)\ ,\\
\nonumber\\
\tr{\rho_A^3} &=&\frac{5+5 \mu+\mu^2}{(1+\mu)^2}\frac{1}{N^2}+O\left(\frac{1}{N^3}\right)\ ,\\
\nonumber\\
\tr{\rho_A^4}&=&\frac{14+21 \mu+9 \mu^2+\mu^3}{(1+\mu)^3}\frac{1}{N^3}+O\left(\frac{1}{N^4}\right)\ ,\\
\nonumber\\
\tr{\rho_A^5}&=&\frac{42+84 \mu+56 \mu^2+14 \mu^3+\mu^4}{(1+\mu)^4}\frac{1}{N^4}+O\left(\frac{1}{N^5}\right)\ .
\label{eq:traces_thermolim}
\end{eqnarray}
The same can be done for the $N$-degree elementary invariant, that is $\det{\rho_A}$. For example, for a balanced bipartition we have
\begin{equation}
\det{{\rho}_A}\sim \frac{N!}{N^{2N}}. 
\label{eq:8}
\end{equation}

\section{Typical states of fixed entanglement}

So far, we have dealt with unbiased pure states. We now look for the most probable eigenvalues sampled on isopurity manifolds, i.e.\ the most probable value of the pdf $f _{N,M}(\bm{\lambda})$, given in~(\ref{eq:Haar_invariant}), on the manifolds
\begin{equation}
\mathcal{M}(\pi_{AB})=\Big\{\bm{\lambda}\in \Delta_{N-1} \, | \, \sum_i{\lambda_i^2}=\pi_{AB}\Big\}\ ,
\label{eq:ROI2}
\end{equation}
which geometrically are given by the intersection of the $N-1$-dimensional sphere of radius $\sqrt{\pi_{AB}}$ with the simplex $\Delta_{N-1}$.
The new constraint, $\sum {\lambda_i^2}= \pi_{AB}$, enables us to compute immediately the second moment of $\lambda_i$, that is, the width of the pdf restricted on the isopurity manifolds. From the permutation invariance of $f_{N,M}$: 
\begin{equation}
\pi_{AB}=\avg{\sum_i{{\lambda_i}^2}}=\sum_i{\avg{{\lambda_i}^2}}=N\avg{{\lambda_i}^2}\quad\Longrightarrow\quad \avg{{\lambda_i}^2}=\frac{\pi_{AB}}{N},\quad\,\, 1\leq i\leq N\ ,
\label{eq:avg_lambdasq}
\end{equation}
and then, the variance is: 
\begin{equation}
\avg{\avg{{\lambda_i}^2}}=\avg{\bigl(\avg{\lambda_i}-\lambda_i\bigr)^2}=\avg{{\lambda_i}^2}-{\avg{\lambda_i}}^2=\frac{1}{N}\Bigl(\pi_{AB}-\frac{1}{N}\Bigr),\quad\,\, 1\leq i\leq N\ ,
\label{eq:sig_lambda_purity_fixed}
\end{equation}
where, in the last equality the constraint on the simplex was used.

By introducing a second Lagrange multiplier $\eta$ to take into account the new constraint, one has to find the minima of the $(N+2)$-variable function
\begin{equation}
\tilde{E}_{\rm{tot}}(\bm{\lambda},\xi,\eta)=-2\sum_{i<j}\log{|\lambda_i-\lambda_j|}-(M-N)\sum_l{\log{\lambda_l}}-\xi(1-\sum_i{\lambda_i}) -\eta(\pi_{AB}-\sum_i{\lambda_i^2}).
\label{eq:landscape_with_2constraint}
\end{equation}
The saddle point equations~(\ref{eq:SaddlePointEqns}) modify into
\begin{equation}
\begin{sistema}
{\displaystyle -2\eta\lambda_i+2\sum_{j\neq i}{\frac{1}{\lambda_i-\lambda_j}}+\frac{M-N}{\lambda_i}-\xi=0\ ,\qquad 1\leq i\leq N,} \\ 
{\displaystyle\sum_i{\lambda_i}=1,}\\
{\displaystyle\sum_i{{\lambda_i}^2}=\pi_{AB}\ .}
\end{sistema}
\label{eq:SaddlePointEqnsBeta}
\end{equation}
As done before, by multiplying the first $N$ equations of system (\ref{eq:SaddlePointEqnsBeta}) by $\lambda_i$ and taking the sum we find
\begin{equation}
\xi=N(M-1)-2\eta\pi_{AB}\ .
\label{eq:lag_rel}
\end{equation}
As pointed out before, to find the most probable spectrum  is analogous to the problem of finding the equilibrium positions of $N$ interacting charges in an external field. The charges repel electrostatically via a $2D$-Coulomb potential in an external potential that now includes a new term due to $\eta\neq0$
\begin{equation}
\varphi_{\rm{ext}}(x)=-\alpha\ln{x}+\xi x+\eta x^2\ ,
\label{eq:ext_field_beta}
\end{equation}
where $\alpha=M-N\geq0$. At the origin there is a charge $\alpha$ that repels the other charges. Moreover, the unit charges experience a constant force whose direction is  opposite to the sign of $\xi$, and the Lagrange multiplier $\eta$ plays the role of the elastic constant of a harmonic potential $\eta x^2$.
Thus, we have reduced the problem  to that of finding, among all configurations of the Coulomb gas with constrained center of mass and momentum of inertia, the one with smallest electrostatic energy. 
Another, equivalent, point of view, is to look at $\eta$ as an inverse  temperature which fixes the energy $\pi_{AB}$ of the system, and will be discussed in section~\ref{sec:canonical}.

As discussed in the previous section, when $\eta=0$ the Lagrange multiplier $\xi$ has to be positive in order to have an energy minimum in $(0,1)$. The minimum is at $x_c=\alpha/\xi$. Since we expect that this critical point is close to $1/N$ we can immediately guess $\xi\sim NM$ in the large $M$ limit. When $\eta\neq0$ the scenario becomes more interesting. Again the $N$  charges will arrange themselves in the external potential in a configuration of minimum energy. In the following analysis, we will consider the balanced situation $\alpha=M-N=0$.

\subsection{Balanced bipartition} 

\label{sec:balanced}

The balanced case, $M=N$, is exactly solvable in the context of orthogonal polynomials. The saddle point equations~(\ref{eq:SaddlePointEqnsBeta}) specialize to 
\begin{equation}
\begin{sistema}
{\displaystyle -2\eta\lambda_i+2\sum_{j\neq i}{\frac{1}{\lambda_i-\lambda_j}}-\xi=0\ ,\qquad 1\leq i\leq N,} \\ 
{\displaystyle\sum_i{\lambda_i}=1,}\\
{\displaystyle\sum_i{\lambda_i^2}=\pi_{AB}\ .}
\end{sistema}
\label{eq:SaddlePointEqnsBeta1}
\end{equation}
By summing over $i$ and by using Eq.(\ref{eq:lag_rel}) with $M=N$ we get
\begin{equation}
\xi=-\frac{2\eta}{N}\ ,\qquad \pi_{AB}=\frac{1}{N}+\frac{N(N-1)}{2\eta}\ .
\label{eq:dirty_trick_balanced}
\end{equation}
The Stieltjes method leads us to the differential equation 
\begin{equation}
g''(x)-(2\eta x+\xi)g'(x)+2\eta N g(x)=0,
\end{equation}
whose polynomial solution is the Hermite polynomial \cite{Szego} $H_N\left(\frac{\xi}{2 \sqrt{\eta}}+\sqrt{\eta} x\right)$. We fix the Lagrange multiplier $\xi=-2\eta/N$ in order to satisfy the unit-trace condition. Then, the most probable eigenvalues are the $N$ solutions of the following polynomial equation: 
\begin{equation}
H_N\left(\sqrt{\eta}\left(x-\frac{1}{N}\right)\right)=0\ .
\label{eq:hermite_beta}
\end{equation}
Moreover, from Eq.\ (\ref{eq:dirty_trick_balanced}), we find how $\eta$ labels different isopurity manifolds
\begin{equation}
\eta=\frac{N^2(N-1)}{2(N\pi_{AB}-1)} \ .
\label{eq:beta_purity}
\end{equation}
Therefore, since $1/N\leq \pi_{AB} \leq 1$, one gets that $\eta \geq N^2/2$, the minimum being attained for separable states, $\pi_{AB}=1$, while $\eta \to +\infty$ for $\pi_{AB} \to 1/N$.

However, if $\eta$ is too small the solution to~(\ref{eq:hermite_beta})
ceases to be physical, since one or more eigenvalues become negative. That means that the most probable spectrum belongs to the boundary of the simplex, and it is no longer a critical point of the total energy.
Thus, for all $N$, there exists a threshold value $\eta_+>N^2/2$ below which the solution suddenly ceases to exist. 
Now, recall that, denoting by $\sigma_N$ the smallest zero of the Hermite polynomial $H_N(x)$, one gets \cite{Szego} 
\begin{equation}
\lim_{N\to+\infty}\frac{\sigma_N}{\sqrt{N}}=-\sqrt{2}\ .
\label{eq:9}
\end{equation}
For large positive values of $\eta$ we expect to obtain very mixed states with $\pi_{AB}=O(1/N)$. Then, in the large $N$ limit, from Eq.\ (\ref{eq:beta_purity}) the correct scaling for the Lagrange multiplier $\eta$ is 
\begin{equation}
\eta=\beta N^3,\qquad \beta \geq 0 .
\label{eq:scaling_beta}
\end{equation}
Using the above asympotic property of the largest zero of the Hermite polynomials, we readily find $N x_N \to 1-\sqrt{2/\beta}$, whence the critical value is $\beta_+=2$, and the corresponding critical values of the purity $\pi^c_{AB}=5/4N$, in agreement with refs.~ \cite{FacchiPascazio,ADePasquale}. The typical eigenvalues of the fixed-purity ensemble are the zeros of Eq.\ (\ref{eq:hermite_beta}), whenever $\pi_{AB}\leq5/4N$.

\section{Canonical Ensemble and Partition Function}

\label{sec:canonical}

The above approach is based on the microcanonical ensemble, in which the purity of distinct manifolds
is fixed. A different way to proceed is to fix the average purity by introducing a partition function and 
reformulate the problem in terms of a classical canonical ensemble. 
The main quantity we are interested in is the  local purity $\pi_{AB}$ as a measure of the bipartite
entanglement between balanced bipartitions. This quantity will play the role of energy
in the statistical mechanical approach. 

Let us clarify the rationale behind our analysis. 
Although our interest is focused on the microcanonical features
of the system, namely on ``isoentangled'' manifolds, we find it
convenient to define a canonical ensemble and a temperature. This
makes the analysis easier to handle and it is based on the  equivalence --largely used   in
the statistical mechanical description of large systems--
between the microcanonical
ensemble (in which energy is fixed) and the canonical ensemble (in
which temperature is fixed).

The inverse temperature $\beta$ is a Lagrange multiplier for the optimization
problem. It is the variable that is naturally conjugate to $\pi_{AB}$:
$\beta$ fixes, with an uncertainty that becomes smaller for a larger
system, the level of the purity of the subset of vectors under
consideration, and thus an isoentangled manifold. The use of a
temperature is a common expedient in minimization problems that can be recast in
terms of classical statistical mechanics. We notice that this approach has been fruitful in several other context such as the analysis of mixed states \cite{lewenstein} and multipartite pure entangled states \cite{classical,classical1,classical2} where entanglement exhibits the phenomenon of frustration \cite{frustration}.

Therefore, in order to study the typical properties of a large bipartite quantum system,
we introduce a partition function from which all thermodynamic quantities can be computed
\begin{equation}
\mathcal{Z}_{AB}=\int_{\Delta_{N-1}}{\exp{\left(-\beta N^3\pi_{AB}\right)\, d\nu(\bm{\lambda})}}\ ,
\label{eq:10}
\end{equation}
where $d\nu(\bm{\lambda})= f_{N,N}(\bm{\lambda}) d \bm{\lambda}$, with pdf~(\ref{eq:Haar_invariant}).
Here, $N^3\pi_{AB}$ plays the role of ``energy''
while $\beta^{-1}$ is a  ``temperature'' that control energy, that is entanglement. The factor $N^3$ is chosen in order to make the energy an extensive quantity, $N^{3}\pi_{AB}=O\left(N^2\right)$, 
since $\avg{\pi_{AB}}=O\left(1/N\right)$.

For large $N$ we look at the maximum of the integrand, that is  the maximum of
\begin{equation}
\tilde{\mathcal{Z}}_{AB}= \int_{\mathbb{R}_+^N} \exp(-N^2 V(\bm{\lambda},\zeta,\beta)) \, d\bm{\lambda},
\label{eq:partitionfunction}
\end{equation}
where
\begin{equation}
\label{eq:potential}
V(\bm{\lambda},\zeta,\beta) = \beta  N \sum_j \lambda_j^2  -  \frac{2}{N^2}\sum_{j<k}\ln |\lambda_j-\lambda_k | + \zeta \big(\sum_k\lambda_k- 1\big).
\end{equation}
As before, the Lagrange multiplier $\zeta$ fixes the normalization constraint and allows one to extend the integration from the simplex $\Delta_{N-1}$ to all positive values $\mathbb{R}_+^N$. Notice that, by setting 
\begin{equation}
\eta = \beta N^3, \qquad  \xi = \zeta N^2,
\end{equation} 
in agreement with Eqs.~(\ref{eq:scaling_beta}) and (\ref{eq:dirty_trick_balanced}), we get
\begin{equation}
N^2 V(\bm{\lambda},\zeta,\beta) = \tilde{E}_{\rm{tot}}(\bm{\lambda},\xi,\eta) + \eta \pi_{AB},
\end{equation}
where $ \tilde{E}_{\rm{tot}}$ is the energy~(\ref{eq:landscape_with_2constraint}) for a balanced bipartition, $N=M$,
thus establishing the equivalence between the two approaches of Sec.~\ref{sec:canonical} and Sec.~\ref{sec:balanced}.
Indeed, in the thermodynamic limit, the main contribution to the integral of the partition function is given by the maximum of its integrand.   The standard way to solve the problem is to apply a saddle point method, i.e. to look for the stationary point of  $V$, that is the stationary point of $\tilde{E}_{\rm{tot}}$, which gives again Eqs.~(\ref{eq:SaddlePointEqnsBeta1}). Notice, however, that in the canonical framework, the last equation of~(\ref{eq:SaddlePointEqnsBeta1})  is not viewed as a constraint, but rather as a relation between temperature and average purity.

Moreover, in a thermodynamic analysis it could be convenient to rewrite the saddle point equations~(\ref{eq:SaddlePointEqnsBeta1})  in the continuous limit. To this end, by recalling that all the eigenvalues $\lambda_k$ are of order $O(1/N)$, we introduce the 
 empirical distribution
\begin{equation}
\sigma(\lambda)= \frac{1}{N} \sum_j \delta(\lambda - N \lambda_j),
\label{eq:densitydef}
\end{equation}
that in the limit of large $N$ can be approximated by a continuous density function.
Obviously, we get
\begin{equation}
\int \sigma(\lambda)\, d\lambda =1,
\label{eq:density1}
\end{equation}
and by making use of~(\ref{eq:densitydef}),  the saddle point equations~(\ref{eq:SaddlePointEqnsBeta1}) 
can be easily rewritten as
\begin{equation}
\begin{sistema}
{\displaystyle \beta \mu +  \fint 
\frac{\sigma(\lambda)}{\lambda-\mu}\, d\lambda +  \frac{\zeta}{2} = 0, } \\ \\
{\displaystyle\int\lambda \, \sigma(\lambda)\, d\lambda=1,}\\ \\
{\displaystyle \int\lambda^2  \sigma(\lambda)\, d\lambda= \tilde{\pi}_{AB}  \ , 
}
\end{sistema}
\label{eq:contin_SPE}
\end{equation}
where $\mu= N\lambda_i$, $\fint$ is the Cauchy principal value integral, and 
\begin{equation}
\tilde{\pi}_{AB} = N \pi_{AB}.
\end{equation}

\begin{figure}[t]
\centering
\includegraphics[width=0.5\columnwidth]{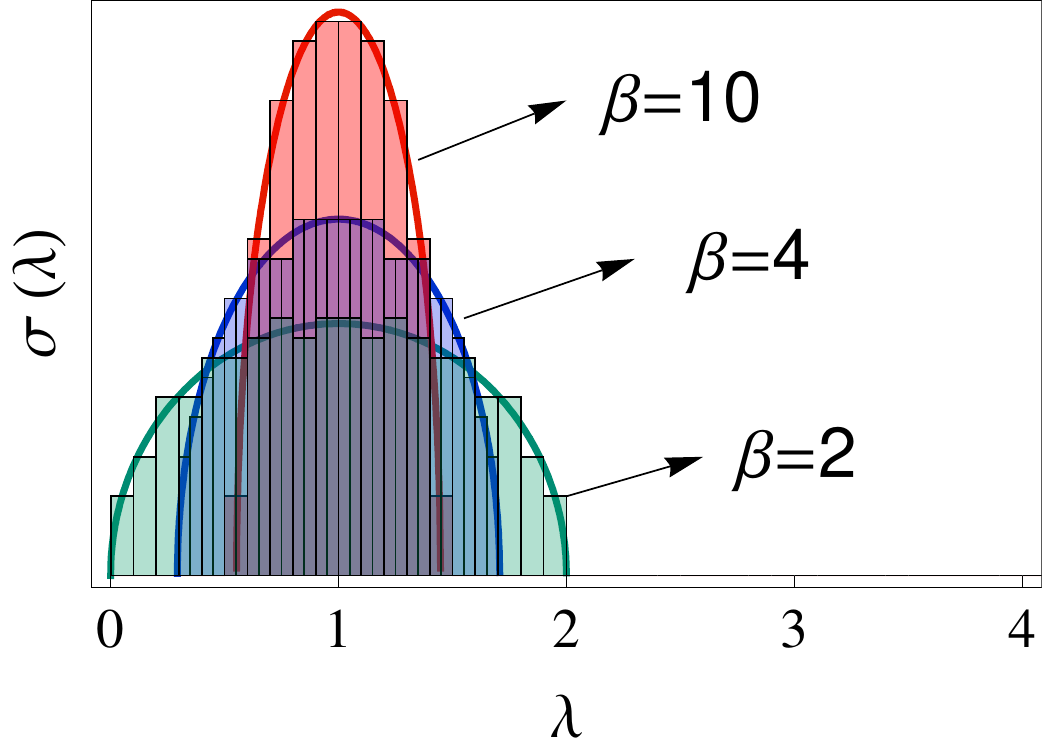}
\qquad \includegraphics[width=0.4\columnwidth]{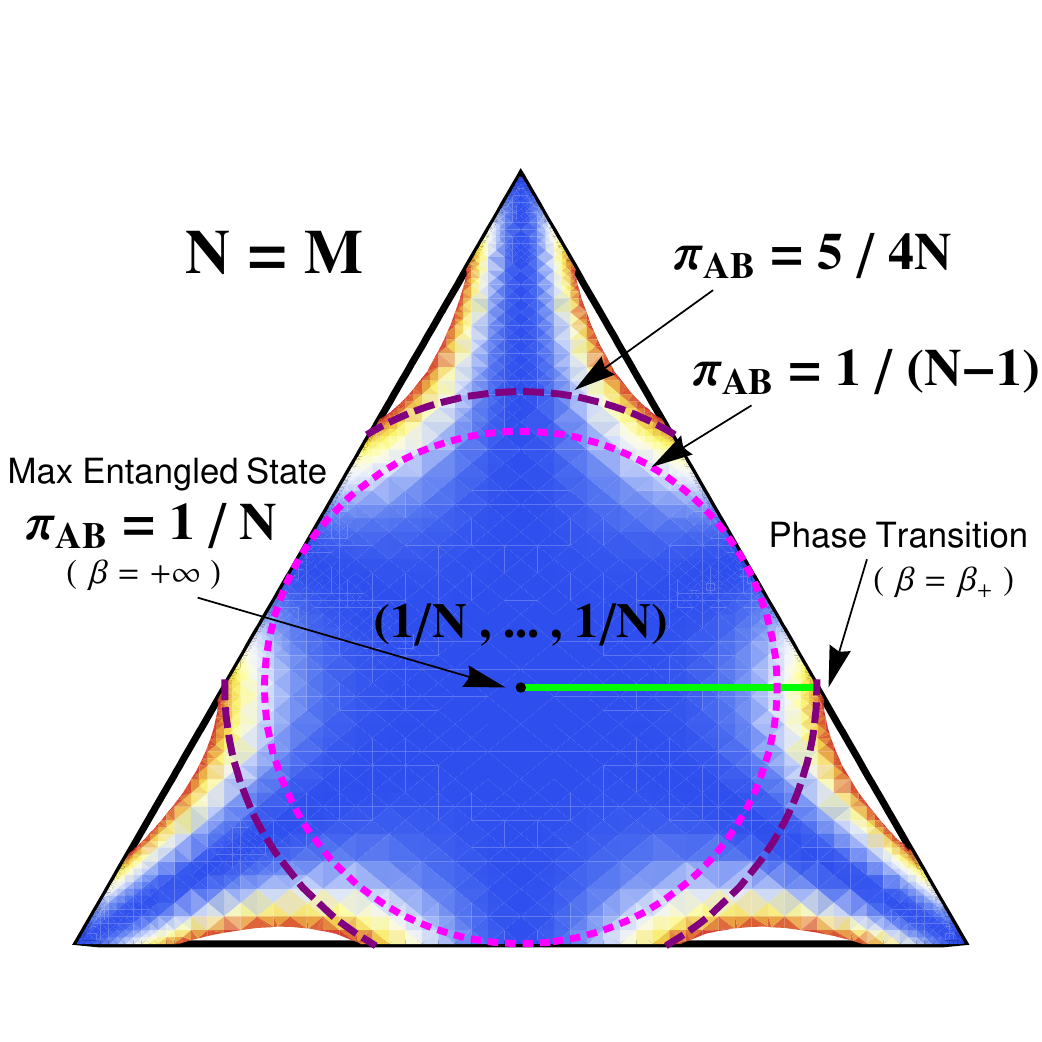}
\caption{Left: Density of the eigenvalues for $\beta = 2,4$, and $10$: analytic solution, Eq.\ (\ref{eq:semicircle}) and numerical results (zeros of Eq.\ (\ref{eq:hermite_beta})). In the temperature range $\beta\in\left[2,+\infty\right)$ the solution is given by the semicircle law. Right: Density plot of the pdf $f_{N,N}(\bm{\lambda})$, Eq.\ (\ref{eq:Haar_invariant}),  in the simplex of eigenvalues of the density matrices obtained by partial tracing in the balanced case $N=M=3$. The green line shows the locus of the most probable spectra (solution of Eq.\ (\ref{eq:hermite_beta})) as $\beta=\eta/N^3$ varies. For $\beta\rightarrow+\infty$ the spectrum tends toward the maximally mixed one $(1/N,\dots,1/N)$, corresponding to a maximally entangled pure state of the total system. As $\beta$ decreases, the typical state becomes less entangled (reduced density matrix less mixed) and reaches the boundary of the simplex for  $\beta=\beta_+$. In the thermodynamic limit, at $\beta_+=2$ a phase transition occurs, one eigenvalue vanishes, and a new equilibrium configuration (Mar\v{c}enko-Pastur distribution) takes place instead of Wigner's semicircle law.}
\label{fig:semicirle}
\end{figure}
The first equation is a singular Fredholm equation of the first kind, known as Tricomi's equation \cite{Tricomi}. 
Its solution lies on a compact interval (depending on $\beta$), and can be given explicitly \cite{Tricomi}. 
One obtains for $\beta\geq \beta_+=2$  Wigner's semicircle law 
\begin{equation}
\label{eq:semicircle}
\sigma(\lambda)=\frac{\beta}{\pi}\sqrt{(\lambda-\lambda_{-})(\lambda_{+}-\lambda)},
\end{equation}
where (see ref.  \cite{ADePasquale})
\begin{equation}
\lambda_{\pm}= 1 \pm\sqrt{\frac{\beta_+}{\beta}},\qquad  \tilde{\pi}_{AB}=1+\frac{1}{2\beta}\ .
\end{equation}
In Fig. \ref{fig:semicirle} we plot $\sigma(\lambda)$ for several values of $\beta$. Observe that as $\beta$ becomes larger the distribution becomes increasingly peaked around $1$. This means that in~(\ref{eq:densitydef}) all the eigenvalues tend to $1/N$: for temperatures $1/\beta$ close to zero the quantum state becomes maximally entangled.

At higher temperatures $0\leq \beta \leq \beta_+$ the solution acquires a different physiognomy, namely the Mar\v{c}enko-Pastur law \cite{MarcenkoPastur}.
The change from semicircle to Mar\v{c}enko-Pastur, a breaking of the $\mathbb{Z}_2$ symmetry of the solution about $1$, is  known  as the Wigner-to-Wishart second order phase transition \cite{FacchiPascazio}. We just mention that, for positive $\beta$'s, lowering the temperature  will make the state nearest to the separable ones. As a consequence, among all the isopurity manifolds the one corresponding  to $\beta=0$, namely the unbiased ensemble, exhibits the highest purity:
\begin{equation}
\sigma(\lambda)=\frac{1}{2\pi}\sqrt{\frac{4-\lambda}{\lambda}},\qquad  \tilde{\pi}_{AB}=2 \ .
\label{eq:13}
\end{equation}

In order to obtain more separable typical states, the sampling has to be done with negative $\beta$'s. By lowering $\beta$ below zero, in the framework of the partition function one finds that $\sigma(\lambda)$ experiences many others phase transitions \cite{ADePasquale}. The analytic continuation of the stable solution for positive temperatures emanates a metastable branch in which all eigenvalues remain $O(1/N)$ (and so does the purity) even though the temperature can be negative. By constrast, a new stable solution emerges by allowing the largest eigenvalue to evaporate, leaving the Mar\v{c}enko-Pastur sea of the other eigenvalues $O(1/N)$ and becoming  of order of unity. This is a signature of separability: negative temperatures explores regions of the pure state space mostly populated by separable states.

\section{Conclusions}
In this paper we have analyzed the properties of the typical spectrum of the reduced density matrices obtained from a suitable ensemble of pure states. We have found that the most probable eigenvalues of the density matrix coincide with the equilibrium positions of charges on a line with interaction forces arising from a logarithmic potential. In particular, for an unbiased ensemble, the solution of the problem can be obtained from the zeroes of Laguerre polynomials.
In the case of a fixed value of purity, the ensemble is no longer unbiased and the charges of the model are affected by a supplementary quadratic term in the external potential. This leads to a modified solution and the most probable eigenvalues are given by the zeroes of  Hermite polynomials.
Finally, in the case of balanced bipartitions, we have used an approach based on a  temperature and a partition function. In this case we have obtained  Wigner's semicircle law, and a second order phase transition to a Mar\v{c}enko-Pastur law.

\section*{Acknowledgments}
P.F.\ and G.F.\ acknowledge support by the project IDEA of Universit\`{a} di Bari.
G.F.\ acknowledges support by Istituto Nazionale di Alta Matematica and Gruppo Nazionale per la Fisica Matematica through the Project Giovani GNFM.
This work is partially supported by PRIN 2010LLKJBX on ``Collective quantum phenomena: from strongly correlated systems to quantum simulators".

%
%
%

%
%
%

\end{document}